\documentclass[useAMS,usenatbib]{mn2e}
\usepackage{epsfig}

\newcommand\mps{$\,{\rm m}\,{\rm s}^{-1}$}

\title[Site testing at Mt.~Shatdzhatmaz]{First results of site testing program at Mt.~Shatdzhatmaz in 2007 -- 2009}
\author[V. Kornilov et al.]{V.~Kornilov,\thanks{E-mail: victor@sai.msu.ru} N.~Shatsky, O.~Voziakova, B.~Safonov, S.~Potanin, M.~Kornilov\\
Sternberg Astronomical Institute, Universitetsky prosp. 13, 119992 Moscow, Russia}

\begin{document} 
\date{Accepted 2010 June 14.  Received 2010 June 14; in original form 2010 May 26}
\pagerange{\pageref{firstpage}--\pageref{lastpage}}
\pubyear{2010}
\maketitle
\label{firstpage}

\begin{abstract}
We present the first results of the site testing performed at Mt.~Shatdzhatmaz at Northern Caucasus, where the new Sternberg astronomical institute 2.5-m telescope will be installed. An automatic site monitor instrumentation and functionality are described together with the methods of measurement of the basic astroclimate and weather parameters. The clear night sky time derived on the basis of 2006 -- 2009 data amounts to 1340 hours per year.

Principle attention is given to the measurement of the optical turbulence altitude distribution which is the most important characteristic affecting optical telescopes performance. For the period from November 2007 to October 2009 more than 85\,000 turbulence profiles were collected using the combined MASS/DIMM instrument. The statistical properties of turbulent atmosphere above the summit are derived and the median values for seeing $\beta_0 = 0.93$~arcsec and free-atmosphere seeing $\beta_{free} = 0.51$~arcsec are determined. Together with the estimations of isoplanatic angle $\theta_0 = 2.07$~arcsec and time constant $\tau_0 = 2.58 \mbox{ ms}$, these are the first representative results obtained for Russian sites which are necessary for development of modern astronomical observation techniques like adaptive optics.

\end{abstract}

\begin{keywords}
site testing -- atmospheric effects -- techniques: miscellaneous
\end{keywords}

\section{Introduction}

Nearly twenty years passed since astronomers recognized that the progress in ground-based optical observations fully depends on the success in minimization of the destructive influence of atmosphere. Indeed, the new generation telescopes have reached maximal technology-limited sizes and diffraction-limited optical quality and the detectors used have attained nearly 100 percent quantum efficiency. The telescope performance was determined then by the last parameter in the Bowen (\citeyear{bowen1964}) formula -- the seeing.

The capabilities of a large telescope may be significantly boosted by the use of adaptive optical systems (AO). Meanwhile, knowledge of solely seeing statistics is insufficient for assessment of the AO perspectives at the given site. The design of AO \citep*[see][]{vern1991, Rodier_book, Wils2003, TPcan, CTIO03} demands the information on the altitude distribution of {\sl optical turbulence} (OT). This need intensified site testing studies about 15 years ago aimed to collect reliable and statistically significant data on OT both in perspective sites \citep*{TMT,EELT,DomeC} and at existing observatories (\citealt{CTIO03}; \citealt*{Palomar,Paranal}).

%REF:Page 1, paragraph 3. This paragraph is grammatically hard to read. It's necessity being questionable, I would simply remove it.
The site selection for the new 2.5-m telescope of Sternberg astronomical institute (SAI, Moscow University) has revealed absence of a reliable and contemporary information on the astroclimate of Caucasian region of Russia. Due to this reason the Shatdzhatmaz summit site near the Kislovodsk city was chosen on the basis of sparse results obtained 40 -- 50 years ago and secondary factors.

%Corr:Page 1, paragraph 4, first sentence should read "astro-climatic" or "astroclimatic"
The night-time astroclimatic characteristics of this place were measured in the end of 1950-s with a star trails photographic technique \citep*{Darchija60,Boshnjak62} which was standard for that time. Authors demonstrated the site advantage over the lower elevation places. Meanwhile, they clearly realised that the average amplitude of stellar image motion $\sigma\sim0.3$~arcsec at zenith related only indirectly to the long exposure image quality.

Such situation has forced us to study the site characteristics with an aim to optimise future observations at the telescope which will be installed there.

The modern way to measure OT is based on the theory of wave propagation through inhomogeneous medium and is focused on determination of the refraction index structural constant $C_n^2$. In typical conditions of night time astronomical observations one can make use of the linear weak perturbations approximation where an atmospheric layer produces the wavefront distortion independently from other layers \citep{Ta67,Roddier81}. This simplification becomes invalid only in unusually strong OT conditions. The applicability of the Kolmogorov model of OT on scales 0.01 -- 3 meters is also verified many times, while in the 3 -- 30\,m domain the von Karman model is normally preferred \citep*{VLT07}.

A number of methods to measure the altitude $C_n^2$ distribution like SCIDAR \citep*{Fuchs}, SLODAR \citep{wilson02}, LOLAS \citep{lolas08}, LUSCI \citep{lusci} was developed which are suited for particular measurement ranges. Meanwhile, the most representative and homogeneous OT data series were accumulated with the DIMM \citep{DIMM} and MASS \citep{MASS} techniques for which one can mention the large-scale site selection programs of the 30-m TMT \citep{TMT} and 42-m E-ELT ESO \citep{EELT} telescopes.

We formulated the main goal of our study as the collection of statistically reliable data on seeing and altitude OT distribution during a 2 -- 3 years campaign. In parallel, the representative information on the amount and quality of clear night sky, on atmospheric transparency, sky background brightness and on-site weather parameters had to be accumulated.

%REF: Page 2, last paragraph of section 1: replace ". Then the automatic site monitor follows" with ", followed by the site..."
In the current paper we describe the developed instrumentation and the results of two-years site monitoring. The detailed data analysis and discussion of the results will be presented later. The next two sections provide a general description of the site selected to build the observatory, followed by the automatic site monitor composition. Section \ref{sec:meteo} presents the two-year statistics of main weather parameters: temperature, wind speed and direction and clear sky data. Afterwards we remind the basics of the MASS and DIMM methods and describe some details of their implementation.

Section \ref{sec:ameba} is dedicated to the observation procedure and describes the measurement campaign performed at Mt.~Shatdzhatmaz. After this we list some data reduction features and particularities. The last section summarises the main OT measurement results and compares them to characteristics of other observatories. Finally the concluding remarks and perspectives are given.

\section{General site description}
\label{sec:gen}
%corr: Page 2, first paragraph of section 2: replace ""being ~50km apart" with "~50km away"
The place to install the 2.5-m telescope is selected near the 2127\,m top of Mt.~Shatdzhatmaz (Karachay-Cherkess Republic of Russia, 20\,km southward from Kislovodsk), approximately 700\,m to south--east from the Highland astronomical (solar) station of Pulkovo observatory (GAS GAO). The mountain belongs to the Skalistiy (`Rocky') ridge which is parallel to the Main Caucasus ridge $\sim$50\,km away. Gradually increasing its height, the Skalistiy ridge approaches Mt.~Bermamyt with a 2642\,m summit some 20\,km to west from our site. In the opposite (SE) direction the ridge declines but has a prominent side extension in the form of a high plateau with the 2880\,m Western Kinzhal top, 18\,km apart from the selected site.

This relief makes the western and northern winds blowing along more or less flat surface while south--eastern winds cross the deep (some hundreds meters) valley of the Hasaut river on the way from Kinzhal to Shatdzhatmaz. These circumstances influence significantly the local climate and atmospheric turbulence characteristics which was also noticed by \citet{gnevkg}.

\begin{figure*}
\begin{tabular}{cc}
\psfig{figure=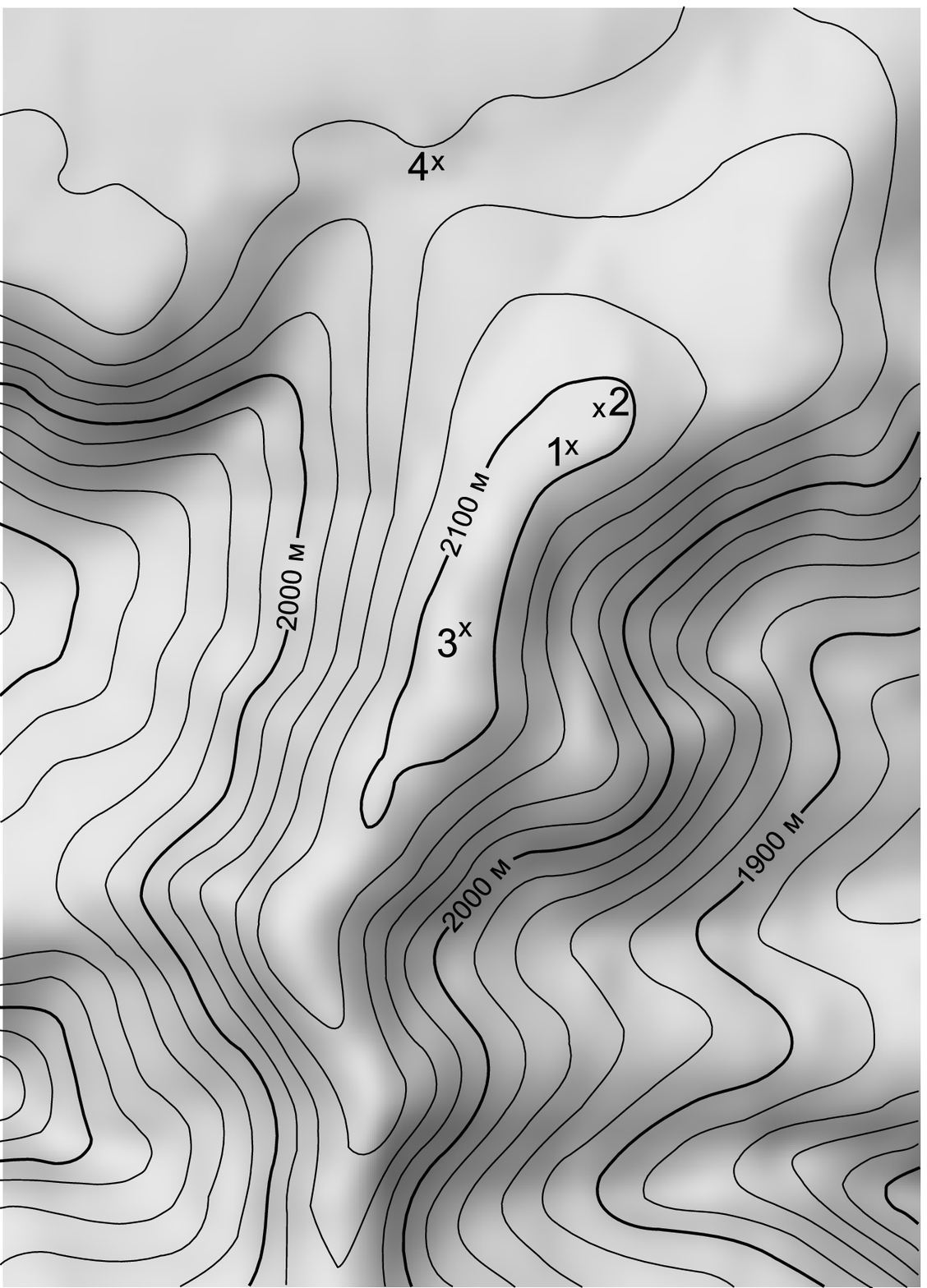,height=6.6cm} &
\psfig{figure=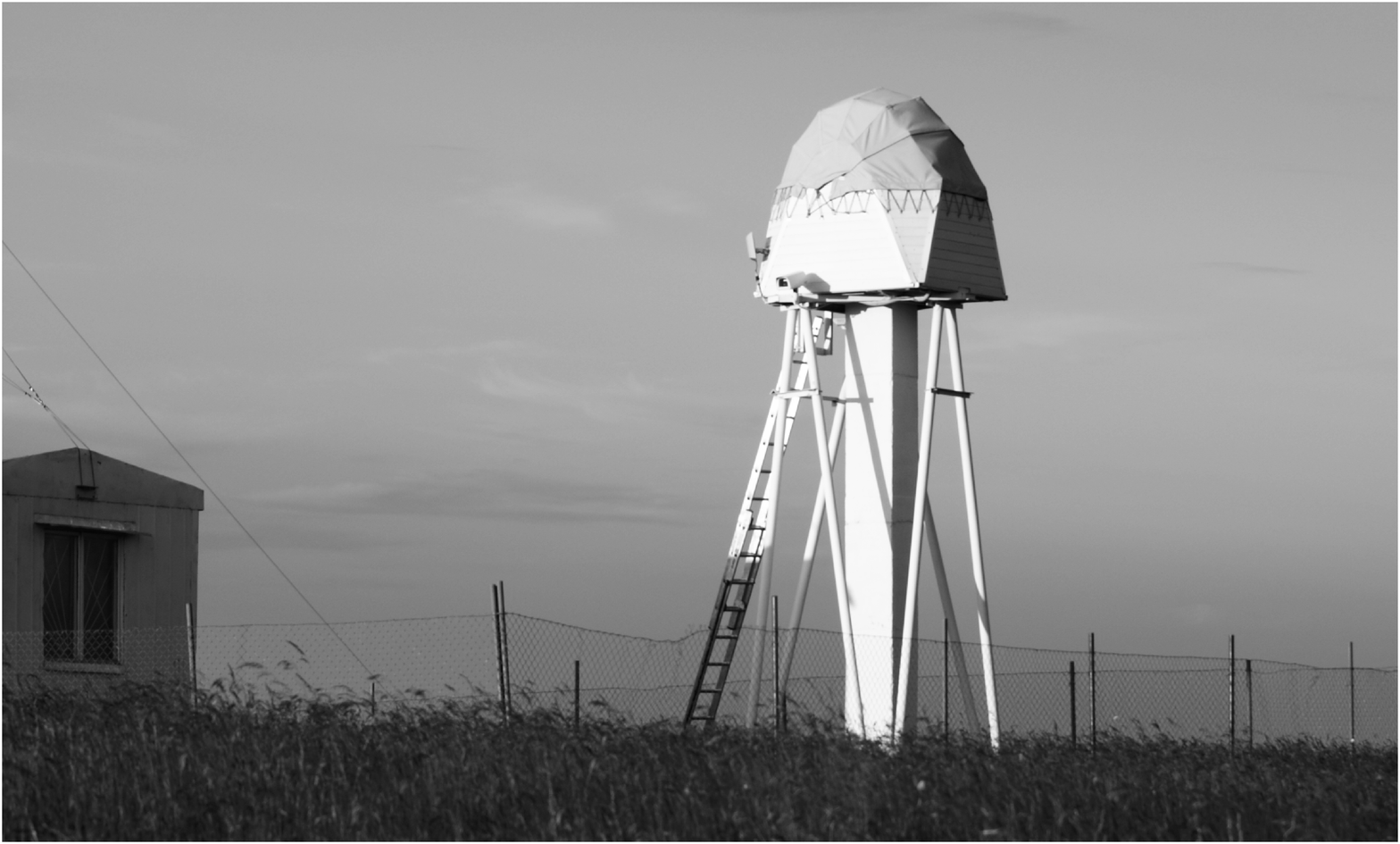,height=6.6cm}\\
\end{tabular}
\caption{Left: the map of the local relief (sides are ca. 2 by 2.5~km). 1 -- automatic site monitor of SAI, 2 -- 2.5-m telescope mark, 3 -- Mt.~Shatdzhatmaz top, 4 -- solar station GAS GAO. Right: the view of the automatic site monitor facility at Mt.~Shatdzhatmaz from the South--West with a lab car to the left. \label{fig:map} \label{fig:view}}
\end{figure*}

The 2.5-m telescope tower will be erected some 40 meters from the steep SE slope of the mountain ($+43^\circ44\arcmin10\arcsec$N, $+42^\circ 40\arcmin03\arcsec$E), at elevation 2112\,m from the sea level (Fig.~\ref{fig:map}). Smaller instruments are planned along the ridge to NE from the 2.5-m telescope tower. The light pollution situation is moderately good: the southern horizon is free from significant sources while there is a glow from towns of Caucasus Mineral Waters region at the northern side. The road Kislovodsk -- Dzhily Su (northern Elbrus) is going half a kilometre to east from the observatory.

Some 100\,km to the west, at northern spurs of the Main ridge, there is Special astrophysical observatory of the Russian Academy of Sciences (SAO RAS), while the Terskol observatory of the Institute of astronomy of RAS is located 50\,km to the south. Unlike the selected site, both observatories are situated close to higher peaks.

Climatic parameters of our site are derived from the data of the meteorological station `Shatdzhatmaz' located 600\,m northward of the 2.5-m telescope tower. The year-averaged temperature equals to $+2.3^\circ$C, the day-time average tempratures vary from $-6^\circ$ in winter to $+12^\circ$ in summer. The mild local climate makes temperatures below $-15^\circ$ and above $+20^\circ$ very rare. The Caucasian humidity is more pronounced in summer (average RH is 80 -- 85~percent), while in winter the weather is frequently dry ($\sim65$~percent). Ground wind is breezy as a rule, although storms exceeding 35\mps\ do happen from time to time. The surface is alpine meadows covered with non-deep snow in winter which provides the low temperature contrast between days and nights. The weather parameters which we collected at the summit (Section \ref{sec:meteo}) are in agreement with these statistics.

\section{Automatic site monitor}

It is well understood that statistically reliable data on OT and other site characteristics are only obtained in course of regular observations during a long (minimum 2 -- 3 years) time span. An optimal solution of this task is to install an automatic site testing monitor functioning with no regular operator intervention. The need of such a robotic facility is supported by both our own experience at Mt.~Maidanak \citep{maid2005} in Uzbekistan and our colleagues' practice.

As a first step, we have developed the suite of automated instrumentation and software for measurement and accumulation of statistics on seeing, altitude OT distribution, clear sky periods, atmospheric transparency and minimal weather parameters (speed and direction of ground wind, temperature and relative humidity). In this development we benefited from experience of similar studies at Cerro Tololo, La Silla, Cerro Paranal and other summits of Chile \citep*{CTIO03,TMTRobot,TMT}.

\subsection{Equipment}

The automatic site monitor (ASM) of SAI was installed at Mt.~Shatdzhatmaz in summer 2007, 40\,m to SW from the spot reserved for the 2.5-m telescope. First remote seeing measurements were conducted in October 2007.

%corr: Page 3, paragraph 2 of section 3.1: "This equipment is powered only during observations due to safety and to minimise power consumption"
The heart of the monitor is the combined MASS/DIMM instrument attached to the automated amateur-class 12-inch telescope Meade RCX400. The telescope and instrument are run by a dedicated computer. This equipment is powered only during observations to assure the detector safety and to minimise power consumption.

The telescope is mounted on a concrete 5-m high pillar having a $0.5\times0.5$\,m section, thus its tube is at 6-m elevation above the ground. Such a height is typical for DIMM facilities at observatories; the 2.5-m telescope primary mirror will also have similar elevation.

The telescope is protected by a custom made enclosure with wooden side walls to minimize thermal effects (Fig.~\ref{fig:view}). The roof (dome) of the enclosure is constituted by steel arches with a tough waterproof fabric pooled on which are moved by a couple of electric actuators. When fully opened, the enclosure allows to reach stars with zenith distances up to $65^\circ$. The enclosure is mounted on the metallic tower composed of three triangles made of 80\,mm pipes which are fixed in 3 individual concrete foundations. Such a rigid construction vibrates quite moderately even at winds exceeding 20\mps. The observation floor is accessed using a lightweight removable ladder.

%corr: Page 3, paragraph 5: replace "provides" with "handles"
Apart from the described OT measurement equipment, there is another compact computer located in the enclosure. It runs days and nights and handles the dome operations, the Wi-Fi link (IEEE 802.11b) with a router at the solar station, the equipment power management and collection of weather data. Two survey webcams are also driven by this machine in order to control the telescope and outdoors area. Meteorological instruments (an anemometer and {\rm T} and {\rm RH} sensors) are fixed at the 5-m height on a special mast atop the lab car roof.

The ASM power is supplied by the storage batteries with a 400~A$\cdot$h total capacity. This provides 7 -- 10 days of operation as the permanently running part consumes about 20\,W and the night-time observations equipment needs 40\,W in addition.

It should be admitted that technical problems affected our site testing during a rather long period, so completely automatic operation began only in February 2008. The major issues were related to the RCX400 scope which is not very suitable for an autonomous work, having a number of drawbacks and some bugs in its firmware. Also, the batteries charging was problematic until the summer 2008, causing interrupts in the observations.

\subsection{Network and software architecture of ASM}

\label{sec:ips}
\begin{figure}
\centering
\psfig{figure=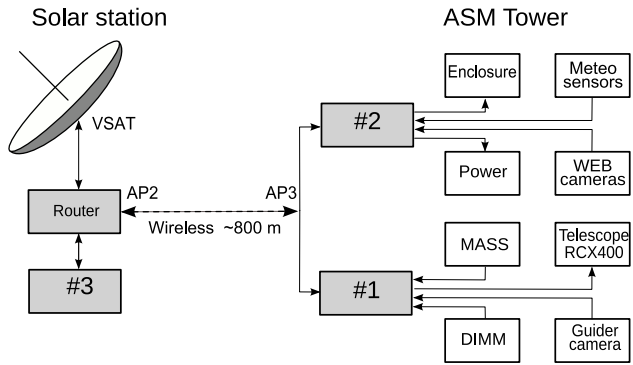,width=8.3cm}
\caption{The network ASM structure. \symbol{35}1 -- OT measurements computer, \symbol{35}2 -- infrastructure control and meteo data acquisition computer, \symbol{35}3 -- observation management, data storage and http-server\label{fig:info}}
\end{figure}

In Fig.~\ref{fig:info} we illustrate schematically the communication structure of ASM. One can see that it is based on three linked computers controlling the hardware. All computers are running free software under the GNU/Linux OS. Data from the custom-made equipment (MASS instrument electronics, meteo controller, dome and power management units) flow via serial RS485 interfaces\footnote{http://curl.sai.msu.ru/mass/download/doc/new\_details.pdf}.

Apart from those described above, there is a computer \symbol{35}3 in the solar station communication room which controls observations and serves as a data storage. Also, it runs an http-server\footnote{URL of ASM server: http://eagle.sai.msu.ru} which allows to control the ASM state via Internet.

The main software consists of a number of independently running applications which communicate with each other via TCP/IP connections. Division of the ASM software into independent components allowed to gain in system flexibility, to maintain the overall system stability and to simplify considerably the logic of each application.
%\footnote{See http://curl.sai.msu.ru/mass/download/doc/infrasoft.pdf [in Russian]}.
The programs are normally running in background as {\sl daemons}.

Each application accomplishes a separate ASM logical task or function. E.g. the {\sc tlsp} program provides telescope pointing, object centring and guiding during measurements for which it communicates with the scope and a finder camera.

The programs are all running as servers, processing only external command requests. An exception is the weather data collecting program {\sc monitor} which starts to poll the meteo controller and to store the results in an output file immediately after start-up. Meanwhile, it also provides the server functions, giving the information on actual conditions when requested.

A simple protocol was implemented\footnote{See http://curl.sai.msu.ru/mass/download/doc/sv\_ug.pdf} for inter-program communication. The commands are given in text (ASCII) form which also allows to manually operate the program components using a system {\sl telnet} utility while debugging or troubleshooting.

In the automatic mode of operation, the programs are controlled by the supervising program {\sc ameba} running on the computer \symbol{35}3 which generates client requests to the programs. It provides logic of the ASM operation according to the algorithms detailed in
Section~\ref{sec:ameba}.

\section{Weather parameters and clear sky statistics}
\label{sec:meteo}
%REF: Page 4, introduction of section 4. Here it would be worth adding the location of the weather sensors. Are they collocated with the DIMM, what is their elevation, etc..
A minimal set of meteo sensors was included in ASM in order to a) provide safe autonomous operation and b) accumulate a homogeneous data set on weather conditions which is also useful for a subsequent analysis of OT monitoring results. No attempt to strictly standardise (calibrate) the measurements was performed. The T, RH and wind sensors are readout by the {\sc monitor} program with a 2-s sampling time to provide instantaneous values which are on-fly averaged in less noisy 1-minute values.

\subsection{Clear night sky}
\begin{figure}
\centering
\psfig{figure=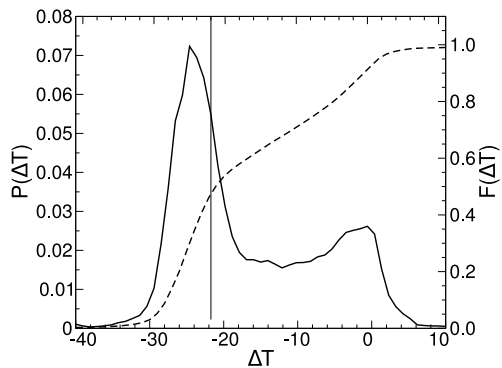,height=6.5cm,angle=-90}
\caption{ Differential (solid line) and integral (dashed line, right axis) distributions of $\Delta T$ in nautical nights.  The thin vertical line depicts the selected threshold used for clear sky detection.
\label{fig:dtemp}}
\end{figure}

%REF: Page 4, beginning of section 4.1: Since they are not standard , can you provide a short explanation of the Boltwood could sensors.
The clear skies were monitored since summer 2006 using the Boltwood clouds sensor\footnote{http://www.cyanogen.com/fix.php} which is connected to a server of the MASTER project \citep{master} at GAS. This bolometric sensor measures infrared radiation of the sky in spectral range from 5 to 10 $\mu$m which gives brightness temperature of the sky. Statistics of clear sky were computed from the 3-years data set spanning the period 2006 November 20 to 2009 October 31. The sky was ranked clear when $\Delta T = T_{sky}-T_{amb}+T_{sen} < -22^\circ$C (where $T_{sky}$, $T_{amb}$ and $T_{sen}$ are temperature readouts from the cloud sensor). This criterion is illustrated by the nautical night time $\Delta T$ distribution in Fig.~\ref{fig:dtemp} where the left peak reflects the clear sky condition.

\begin{figure}
\centering
\psfig{figure=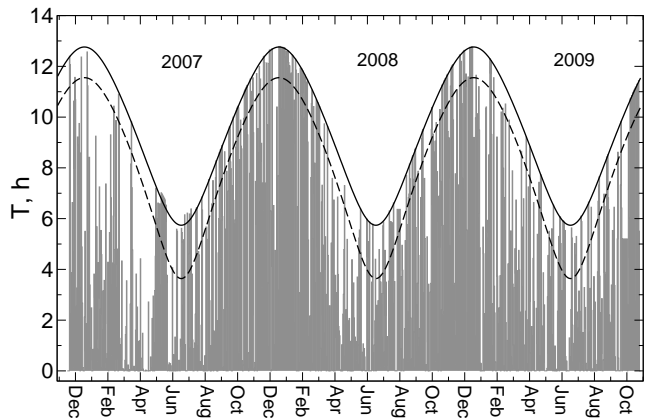,height=8.4cm,angle=-90}
\caption{ Clear sky nights distribution 2006 November 20 to 2009 October 31. The solid line denotes the nautical night duration, the dashed line does for the astronomical night.
\label{fig:clear}}
\end{figure}

Since there are astronomical observations which are not critical to the sky background we expanded the clear sky analysis by two conditions: for astronomical night ($h_{\odot}< -18^\circ$) and for nautical night ($h_{\odot}< -12^\circ$). In Fig.~\ref{fig:clear} the diagram of the clear nautical night sky distribution is shown for the above specified 3-years period.

The annual amount of clear astronomical night sky at Mt.~Shatdzhatmaz is 1343\,h, to be compared to the theoretical total 2950 night-time hours at the given latitude. This amount is slightly higher than the \cite{Rylkov} estimate of 1200\,h. For nautical nights, the annual amount is 1546\,h, while the calculated total is 3451\,h.

The monthly clear-sky astronomical and nautical time distribution is given in Table~\ref{tab:month_clear}. It shows that the most favourable observations period is from mid-September to mid-March, containing $\approx 70$~percent of the yearly total clear sky hours. This period coincides with the time of the slowest high-altitude wind.

\begin{table}
\caption{Monthly amounts of astronomical and nautical clear night sky $CS$ (in hours) and their fractions of the whole night time (in percent)
\label{tab:month_clear}}
\centering
\begin{tabular}{l|rlrl}
\hline\hline
          & \multicolumn{2}{c}{Astronomical} & \multicolumn{2}{c}{Nautical} \\
Months    & $CS$    & Frac. & $CS$ & Frac. \\
\hline
Jan  & 158  & 45   & 175   & 46  \\
Feb  & 131  & 44   & 145   & 46  \\
Mar  & 85   & 31   & 95    & 31  \\
Apr  & 45   & 22   & 53    & 21  \\
May  & 62   & 40   & 80    & 38  \\
Jun  & 44   & 39   & 64    & 36  \\
Jul  & 58   & 42   & 82    & 42  \\
Aug  & 79   & 40   & 94    & 39  \\
Sep  & 101  & 41   & 116   & 41  \\
Oct  & 189  & 62   & 210   & 62  \\
Nov  & 197  & 60   & 216   & 60  \\
Dec  & 193  & 54   & 213   & 54  \\
\hline\hline
\end{tabular}
\end{table}

The direct  comparison of clear  night hours between  observatories is complicated by differing clear-sky criteria and the way the data are presented in terms  of fraction of the whole  night time. However, for Maidanak, La Silla and Paranal observatories the annual clear-sky night time (Table~\ref{tab:clrt})  is recalculated from the data of \cite{eso}. The estimation of the SAO clear skies is reported by \cite{SAO2}. We derived  the clear sky for peak Terskol using the 0 -- 2 cloudiness data from the  paper by \cite*{Ter1}. In the latter study, the authors give the  yearly clear sky duration of 1200\,h, including partially clear weather.

\begin{table}
\caption{Annual clear astronomical night sky at six observatories: actual clear sky (in hours), its fraction of the whole night time (in percent) and the whole night time (in hours)\label{tab:clrt}}
\centering
\begin{tabular}{@{}l|rrr}
\hline\hline
                & Clear sky    & Fraction   & Whole \\
\hline
Shatdzhatmaz    &   1343    &    46 & 2950  \\
SAO             &   1020    &    34 & 2961  \\
Terskol         &    874    &    29 & 2975  \\
Maidanak        &   1697    &    55 & 3088  \\
La Silla        &   1861    &    57 & 3291  \\
Paranal         &   2604    &    78 & 3350  \\
\hline\hline
\end{tabular}
\end{table}

The $\Delta T$ criterion given above is rather strict; it was chosen to avoid frequent ASM shut-downs due to lowering of measured star counts. Such a threshold corresponds rather to the conditions which other authors refer to as `photometric skies'. Softening the threshold to, say, $\Delta T = -18^\circ$C (corresponding to the cloudiness $\approx 25$~percent) boosts the clear night fraction from 46 to 58~percent.

%REF: Page 5, Fig4:On both axis, the \Delta show as crossed circles.

\begin{figure}
\centering
\psfig{figure=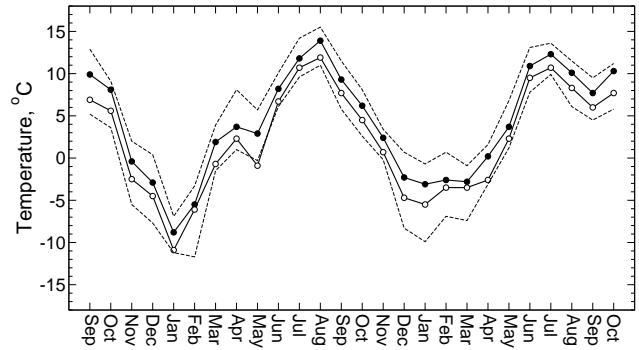,height=8.3cm,angle=-90}
\caption{Two-year evolution of the monthly median day-time (black circles) and night-time (open circles) temperatures. Dashed lines trace the first and third quartiles of monthly distributions.
\label{fig:mtemp}}
\end{figure}

\subsection{Temperature}

The outside temperature sensor is based on the TMP36 circuit (Analog Devices) and is installed in a standard meteorological housing in order to minimize the heating by the sun. The comparison of its reading with data from the cloud sensor and the meteo station records showed the discrepancies to be below $\pm2^\circ$C.

Although our weather data accumulation was started on July 19, 2007, the analysis encompasses the period starting from 2007 November~1. The absolute range of temperatures on clear nights is from $+17.8^\circ$C to $-17.2^\circ$C. The median value is $+1.8^\circ$C which is more typical for observatories at higher altitudes.

Low temperatures in clear nights reduce the typical precipitable water vapour. Obtained from processing of the GPS signal wet delay data, the water content shows the median value of 7.75\,mm. The detailed study on the precipitable water vapour and optical extinction above the summit will be published elsewhere.

As a comparison, the yearly average temperature at SAO which we derived using the weather page data on the observatory website\footnote{http://www.sao.ru/tb/tcs/meteo/data} is $+4.2^\circ$C over the same period. When weighted by the observation time distribution, it might be even higher since the summer -- autumn is the favoured observing season there. At the Terskol site, the yearly average temperature weighted by observing time is $-6.1^\circ$C.

In Fig.~\ref{fig:mtemp} we present the diagram of night- and day-time temperatures averaged within months. The first and third quartiles Q1 and Q3 of monthly distributions are plotted for reference. It is clear that  the two-year data span is not long enough for obtaining statistically reliable monthly averages.

The local turbulence is largely dependent on the contrast between the day- and night-time temperatures. At out site, the diurnal temperature difference is small and amounts to $1.3^\circ$C if computed as the difference of temperatures 1\,h before and 1\,h after sunset. If one takes the full amplitude of daily variation between noon and midnight then it becomes $\approx3.4^\circ$C for those days when the following night was clear.

Similar diurnal temperature variation is observed at SAO RAS where the surface character is similar. At Terskol, the average difference is $\approx2.9^\circ$C as quoted by \cite{Ter1}. These contrasts were computed for the entire data set, without regard to cloudiness.

\begin{figure}
\centering
\psfig{figure=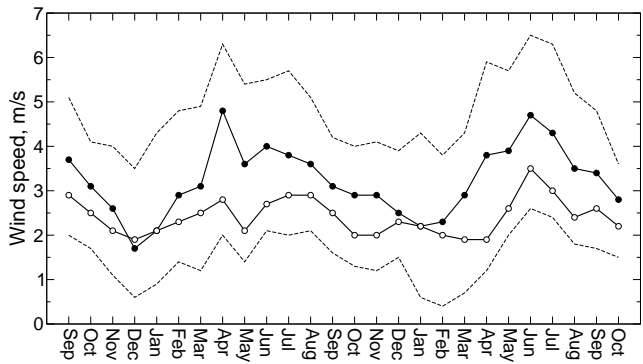,height=8.4cm,angle=-90}\\
\caption{ Two-year evolution of the monthly median wind speed during  daytine (black circles) and night (open circles). Dashed lines  depict the first and third quartiles of monthly distributions.
\label{fig:mwind}}
\end{figure}

\subsection{Ground wind speed}

The strength and direction of ground winds define the airflow conditions about the local relief and the observatory and influences the generation of the ground-layer turbulence.

Wind speed and direction were measured using a conventional cup anemometer by Davis Instruments (model 7911) attached to the meteo controller. Our data show that the wind at the summit is mostly a light breeze, in agreement with the official meteo statistics. In Fig.~\ref{fig:mwind} the monthly median values of the night- and day-time wind speed are shown accompanied by respective quartiles of the distributions. It is noticeable that the night wind is more stable throughout a year while the daytime wind is stronger in summer.

\begin{figure}
\psfig{figure=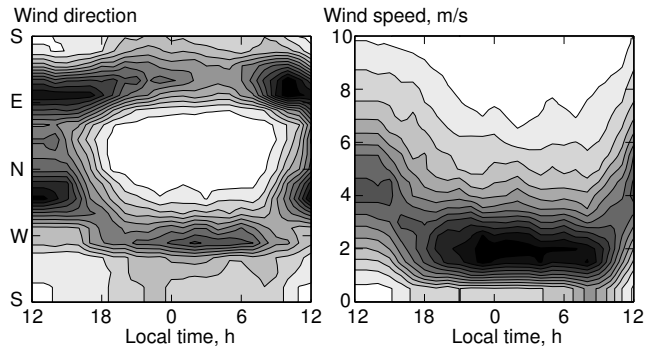,width=8.4cm}
\caption{ Two-dimensional probability distribution of the direction (left) and strength (right) of wind as function of the local time. Midnight is in the middle, more dense (darker) regions correspond to higher probability.
\label{fig:wdir}}
\end{figure}

The global median wind speed is 3.3\mps\ in daytime and 2.3\mps\ at night, both in clear and cloudy time. Similar values are characteristic for the Maidanak summit \citep{eso} which was distinguished as an advantage over many other observatories. Meanwhile, in 1~percent of cases the wind is stronger than 12\mps. In the night time the strong wind is rare; wind speed in excess of 9\mps\ occurs in 2.2~percent of cases. The dependence of wind speed on the time of day is shown in Fig.~\ref{fig:wdir}.

The dominant wind direction at night is mostly from the West and, to a smaller extent, from the South--East. The midday behaviour is different, as the NW winds become more frequent. The probability distribution of the wind direction as a function of time of day is shown in Fig.~\ref{fig:wdir} (left). Apparently, preferable wind directions are related to the large-scale relief properties.

%% really two-dimensional?

The NCEP/NCAR\footnote{http://www.esrl.noaa.gov/psd/data/reanalysis/reanalysis.shtml} is a recognized standard source of the high-altitude wind data. According to it, most of the year (October to May) the wind speed at 13\,km altitude is about 20\mps\ while in summer (June -- September) it increases to 35\mps. This can be expected since the Shatdzhatmaz latitude corresponds to the minimum between the tropical and polar streams. The situation is similar at other moderate-latitude observatories \citep*{Carrasco}.

For comparison, analysis of the weather data at SAO for the 2007 -- 2009 period shows the median/mean night-time wind speed to be 2.0/3.8\mps\ in winter and 1.3/2.3\mps\ in summer. Meanwhile, the probability of strong ($>$9\mps) wind is 0.13 and 0.04, respectively. \cite{SAO1} report slightly different situation with 5.9\mps\ in October -- March and 2.5\mps\ in April -- September recorded in the 1960-s. Similarly to SAO, the winds at peak Terskol are stronger in winter than in summer: 2.5 -- 3.6\mps\ and 1.4 -- 2.6\mps, respectively.

\section{Optical turbulence measurement}

\subsection{The principles of MASS and DIMM instruments}

Let us recall that the OT power is described by a single parameter in the Kolmogorov model. This is normally the Fried radius $r_0$ (atmospheric coherence radius). Meanwhile, a more appropriate measure for description of the turbulence distribution along the line of sight is the turbulence intensity $J$ (having the dimension of $\mbox{m}^{1/3}$) which is defined as
\begin{equation}
J = \int C_n^2(z)\, dz,
\label{eq:Jdef}
\end{equation}
where the integration is performed along the whole light path. $J$ is related to the Fried radius as:
\begin{equation}
J = 0.06\, \lambda^2 r_0^{-5/3} \quad\mbox{or}
\end{equation}
\begin{equation}
J = 1.5\cdot10^{-14}\, r_0^{-5/3}, \quad\mbox{if } \lambda = 500\mbox{ nm}
\label{eq:Jr0}
\end{equation}
Unlike other parameters, $J$ is additive and does not depend on wavelength. In order to compare the measurements performed at different air masses $M$, one should reduce the measured $J'$ to zenith, where the light propagation distance $z$ is replaced by height $h$: $J = J'/M$.

It is more common to quantify the full turbulence at a given site by the image quality $\beta$ (seeing) which has a sense of a characteristic size of the turbulent disc (full width at half maximum -- FWHM) registered with a long exposure in a large ideal telescope. For $\lambda = 500\mbox{ nm}$ this measure (in arcseconds) is
\begin{equation}
\beta = 2\cdot10^7\, J^{3/5}
\label{eq:beta}
\end{equation}

The theory of the DIMM instrument (Differential Image Motion Monitor) is described by \cite{DIMM} and \cite{T2002}. The principle is based on measurement of differential displacement of two star images constructed by two circular apertures. These apertures have the diameter $D$ about 10\,cm each and are separated by the baseline distance $d$ about 20\,cm in the telescope entrance pupil plane. The star images are offset from each other by additional optical elements.

Such a differential motion is described in terms of differential {\sl slopes} of respective wavefront regions, related to the OT power. The variance of the longitudinal differential motion $\sigma^2_l$ (along the apertures baseline) and of the transversal motion $\sigma^2_t$ (perpendicular to the baseline) is expressed as
\begin{equation}
\sigma^2_{l,t} = 16.7\cdot K_{l,t}\;D^{-1/3} J.
\label{eq:KDIMM}
\end{equation}
The equation (\ref{eq:KDIMM}) is obtained from a conventional expression from above cited papers by the substitution of $r_0$ from the formula (\ref{eq:Jr0}).

For the Kolmogorov's turbulence spectrum, the coefficients $K_{l,t}$ depend solely on the apertures geometry \citep{DIMM}:
\begin{equation}
 \begin{array}{l}
K_{l} = 0.358 \, (1 - 0.541 \, (d/D)^{-1/3}) \\
K_{t} = 0.358 \, (1 - 0.811 \, (d/D)^{-1/3}). \\
\end{array}
\label{eq:K}
\end{equation}
Thus the DIMM sensor allows to measure the full OT intensity from the telescope entrance to the upper boundary of the atmosphere. This is an essential but still insufficient information.

Multi Aperture Scintillation Sensor (MASS, \citealt{MASS}) is a technique allowing to obtain the detailed information on the OT {\sl vertical distribution} This method is based on simultaneous measurements of stellar scintillation in four concentric entrance apertures. This allows to obtain 4 normal and 6 differential scintillation indices $s^2$ which are the variances of relative light intensity fluctuations. Scintillation is measured with 1-ms exposure and $s^2$ are computed for 1-s time intervals \citep[][hereafter TKSV]{MASS,mnras2003}.

The weak perturbation theory implies that scintillation indices $s^2$ are produced by the whole atmosphere in the following way:
\begin{equation}
s^2 = \int_0^\infty C_n^2(z) Q(z)\,dz,
\label{eq:sc}
\end{equation}
where $Q(z)$ is a weighting function depending on the wave propagation geometry: the entrance apertures dimensions and angular size of a source. For the particular instrument geometry and assumed spatial spectrum of scintillation the weighting function can be computed theoretically (\citealt{T2002time}; TKSV).

Having 10 equations like (\ref{eq:sc}) with different weighting functions, one can solve the inverse problem and obtain the OT profile from the measured indices $s^2$. For a feasible set of entrance apertures (defining the respective set of $Q(z)$) the problem is ill-conditioned so the non-trivial procedure of the vertical profile restoration constitutes a key part of the MASS technique (TKSV; \citealt{exp2010}). The main drawback of the scintillation method is its blindness to the ground layer turbulence, so without additional measurements in this important domain the results can suffer from considerable incompleteness.

\subsection{The combined MASS/DIMM instrument}

Since the MASS apertures can be inscribed in a $\sim\!10$\,cm circle, both MASS and DIMM methods may be implemented in a single instrument fed by a 10 -- 12-inch sized telescope. This was realized in late 2003 and the details of the construction of this combined device and its properties are discussed by \cite{MD2007}. The advantages of this approach are obvious: one needs a single telescope, a DIMM channel camera serves for centring and guiding of a star, both methods deal with the same volume of turbulent atmosphere so its non-stationary nature does not prevent the joint analysis of data. These devices were involved in a majority of large-scale site testing campaigns of the last years \citep{TMT,EELT}.

Our instrument differs by the model of the DIMM channel detector. A high-speed industrial CCD camera EC650 from Prosilica with Firewire interface (IEEE1394) is used. The camera fully complies with the IIDC standard, so the open library {\sc libdc1394}\footnote{http://damien.douxchamps.net/ieee1394/libdc1394/} may be used to program the data exchange under GNU/Linux OS.

Setting the exposure time to 4\,ms allows to achieve the image rate of $\approx 200$ frames per second in the region-of-interest (ROI) $60\times100\mbox{ pixels}$ -- much faster than with other cameras
used in DIMMs. Such a frame rate allows to measure the variances $\sigma^2_{l,t}$ with a maximal possible accuracy. The low readout noise of $\approx10e^-$ allows to use stars up to $4$ magnitude.

An important calibration parameter of DIMM is the pixel scale of the detector since $\sigma_{l,t}$ in eq.(\ref{eq:KDIMM}) must be in radians. Initially we estimated the scale in a conventional way with help of the visual double star HR7141/2 = $\theta$~Ser having a $22.3$~arcsec separation. The diurnal sky rotation appeared to provide us with a more convenient and precise way to calibrate the scale delivering the value 0.612~arcsec per pixel. The entrance aperture dimensions $d$ and $D$ are set by the physical sizes of the DIMM mask which is located in the exit pupil of the `telescope+instrument' system and by the magnification $k$ of this system. The $d/D$ ratio was measured as $2.18\pm0.02$, the aperture sizes $D$ were 90\,mm both.

The OT profile restoration is based on precise knowledge of weight functions (\ref{eq:sc}) which also depend on the system magnification $k$ determining the effective sizes of entrance apertures. So, each focusing of the Fabry lens was followed by the measurement of $k$ with 1-2~percent precision ($k = 16.3$ in the final state). Since functions $Q(z)$ involve also chromatic effects and thus depend on the spectral response of MASS detectors, the latter was controlled\footnote{ http://curl.sai.msu.ru/mass/download/doc/\\ \rule{1cm}{0pt}mass\_spectral\_band\_eng.pdf} by special observations of stars with a wide range of colours similarly to the technique of \cite{maid2005}.

The MASS channel data reduction requires also other instrumental parameters, namely the detectors non-poissonity $p$ and non-linearity $\tau$ which serve to correct exactly for the photon-noise contribution to measured variances \citep{MASS}. These parameters were also monitored using dedicated measurements.

\subsection{MASS/DIMM control and data acquisition}

Unlike the service applications described in Section~\ref{sec:ips}, the MASS/DIMM driving programs {\sc mass} and {\sc dimm} affect not only the overall ASM efficiency but the reliability of the results themselves.

%corr: Page 8, second paragraph: "the absence"
The principal functionalities of the {\sc mass} program are practically the same as those of preceding {\sc Turbina} program and are described in (TKSV; \citealt{MD2007}). The main difference is the absence of graphical user interface and real-time profiles restoration which proved to be a source of instability in automatic and remote mode of operation.

The program performs continuous light intensity recording in A, B, C and D apertures with elementary integration time of 1\,ms during 1-s basic measurement. This data set is on-line processed in order to
compute different statistics: variances and covariances with time lag 1 and 2 for normalized signals and their combinations. These statistics are stored in the output file for subsequent processing.

Measurements are started by an external command and repeated over an accumulation time. The mentioned integration, basic and accumulation times along with other parameters specifying the measurement process are read by the program from a configuration file during initialization. No command can change these settings which ensures the good homogeneity of output results.

The {\sc dimm} program for communication with the DIMM-channel camera has a similar functional structure. Each subframe obtained from the video stream (containing the image of the ROI) is processed to recognize two stellar images for which the gravity centres and other characteristics are computed. The frame exposure, measurement time and accumulation time are also defined by a configuration file.

A special command initiates the measurement of the star position in a full frame when searching for the new target and starting guiding.

\section{Observing campaign}
\label{sec:ameba}

\subsection{The RCX400 telescope peculiarities}

Remote and automatic observations demand pointing accuracy of about 1~arcmin in order to get a target well inside the MASS/DIMM field diaphragm having a $\approx 2$~arcmin radius. Such an accuracy is achievable with RCX400 after one determines the telescope coordinate system using the 3-stars method. Meanwhile, bugs in the instrument firmware cause the telescope controller to occasionally loose this reference, while its remote recovery does not provide the needed precision.

To cope with this situation, we attached a web camera (Philips SPC900NC) to the telescope finder which allows to detect stars up to $8$ magnitude in the $0.9^\circ\times1.2^\circ$ field of view. The {\sc tlsp} program uses images from that camera to search, centre and guide targets. But the accurate guiding during OT measurements is made using the DIMM channel information.

Introduction of a CCD-finder and development of optimized procedures to restore the telescope coordinate system after powering it on and to explicitly control the axes movement while slewing solved the pointing
problem. More than 3500 successful pointings were performed during the campaign.

Another major RCX400 problem is its focusing which gradually deteriorates the optical adjustment (so-called collimation). The accumulating tilt errors of the secondary mirror produce coma which is seen as astigmatism in the DIMM apertures, biasing measurements of OT strength \citep{TK07}. Meanwhile, the documented collimation procedure can only be performed interactively. Only in the autumn 2009 we overcame this restriction\footnote{See http://curl.sai.msu.ru/mass/download/doc/rcx\_advices.pdf}.

In general, the `telescope+DIMM' optical system was maintained in a good state. The quality was controlled by the Strehl ratio $S$ of the DIMM images; its median value over the whole campaign was 0.47 for both
apertures\footnote{http://curl.sai.msu.ru/mass/download/doc/shtrel\_controls.pdf}. Only in the period from April to June 2009 the telescope alignment was worse and $S$ progressively decreased to 0.3 (left image) and 0.2 (right image). At most 5~percent of measurements were performed in this state.

Another annoyance was dew condensation on the telescope corrector plate in windy and humid conditions. Invention of a special-shape dew cap in November 2007 allowed us to get rid of this problem.

An additional factor limiting the measurement ability is a wind-caused telescope vibration. In slow wind conditions the common image motion is about $\pm0.5$~arcsec and is mainly caused by cophased turbulence
effects and telescope tracking errors. With the wind speed in excess of 3\mps\ the vibration amplitude grows almost linearly and approaches $10$~arcsec in elevation when the wind is 7 -- 8\mps. The amplitude in azimuth is two times less. The wind gusts excite several telescope eigenmodes near 10~Hz, so the vibration amplitude may exceed eventually $20$~arcsec. In this case the ROI must be enlarged and the framerate becomes slower.

More importantly, in these conditions the images are moving so quickly ($>$600~arcsec\,s${}^{-1}$) that the respective smearing exceeds $2.4$~arcsec during exposure. As in the optical aberrations case, this effect introduces DIMM measurement errors and biases the OT estimations up for strong high turbulence. Luckily, 97~percent of measurements were made with vibration amplitudes below $6.4$~arcsec and 85~percent with $<$3~arcsec.

\subsection{Operation control and time coverage}

Control of the ASM is performed by the {\sc ameba} program which is started daily in the evening at the computer \symbol{35}3 and communicates with respective servers.

%REF: Page 8, section 6.2, how did the system know to shutdown during rain?
OT measurement is initiated when the following conditions are fulfilled: 1) the sun altitude is below $-12^\circ$; 2) the cloud sensor reports clear sky; 3) the wind speed does not exceed some limit; 4) the battery voltage is normal. Humidity was not decisive because of the odd sensor behaviour near the dew point, when condensing moisture induced an abrupt lowering of the sensor reading.

\begin{figure}
\centering
\psfig{figure=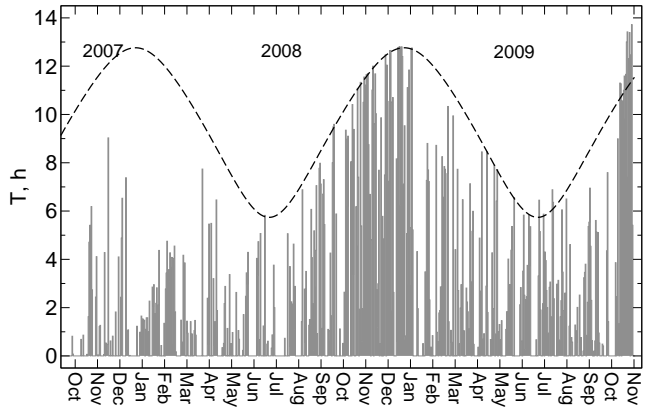,height=8.4cm,angle=-90}
\caption{ Duration odf measurement plotted for the whole period of ASM monitoring. Dashed line is the nautical night duration ($h_{\odot} < -12^\circ$).
\label{fig:duty}}
\end{figure}

When all conditions are satisfied, an observation session is started consisting of the following stages:
\begin{enumerate}
\item Dome opening, powering up the observation equipment, {\sc tlsp, dimm} and {\sc mass} programs initiation.
\item Target selection from a list\footnote{http://curl.sai.msu.ru/mass/download/doc/new\_mass\_cat.pdf} of 119 bright stars according to its magnitude, altitude, distance to the Moon and available measurement duration.
\item Pointing to the target followed by its search, centring, brightness check and guiding start-up.
\item Background measurement in the MASS channel is performed before and after target measurements.
\item In the case of the failed pointing or high background, another target is selected.
\item The target measurement session is performed during 1.5\,hours or until one of the selection criteria is violated.
\item In the case of an error during measurement (equipment failure, stellar flux drop) another target is selected.
\item If an error occurs three times in row, the session is finished; the dome is closed and the equipment is switched off.
\end{enumerate}

After the session is finished by worsened conditions, {\sc ameba} waits for their restoration. In order to prevent random effects, the hysteresis is introduced. For example, observations stopped by the wind speed in excess of 9\mps\ will be resumed after the wind speed remains below 6\mps\ during 20 minutes. Such a pause is obligatory after any session is stopped by an error. Note that the system initialization and parking lasts 2 -- 3 minutes. At the end of the night, {\sc ameba} stops conditions checking or a measurement session and shuts itself down.

%Algorithms of automatic ASM operation were developed till February 2008; before that, observations were made under remote control. Such observations gained us 10~percent of data. %%%Automatic operation increased the data rate by 10~percent.
In the course of a few months after the campaign beginning we have been operating ASM remotely while simultaneously developing and testing the automatic operation algorithms. This period supplements 10 percent of the data. Fully automated observations were started in February 2008.

In Fig.~\ref{fig:duty} the duration of nightly measurement sessions is plotted for every calendar date of the whole campaign. The total observation duration defined as the sum of time spans between the dome was opened and closed is 1753\,h. The net measurement time is 1485\,h or $\approx 90\,000$ OT determinations, while about 18~percent of time was spent for service operations: pointing, centring, background measurements etc. In October 2009 the observation sessions lasted longer than nautical nights since experimental twilight observations were introduced.

In Table~\ref{tab:obs} we list the total monthly measurement duration for the whole period of the campaign. One should beware that nights were accounted as observational when even a single successful OT measurement was performed.

\begin{table}
\caption{Monthly duration of measurements in nights (n) and hours (h) in 2007 -- 2009. Efficiency (E) is given in percent.\label{tab:obs}}
\begin{tabular}{l|rrr|rrr|rrr}
\hline\hline
Year &\multicolumn{3}{c}{2007} & \multicolumn{3}{|c|}{2008} & \multicolumn{3}{c}{2009}\\
Month & n & h   & E & n   & h  & E & n & h & E \\
\hline
Jan  & --  & -- &-- & 19 &36  & 12 & 14  & 63  & 44 \\
Feb  & --  & -- &-- & 21 &57  & 26 & 17  & 83  & 79 \\
Mar  & --  & -- &-- & 11 &24  & 18 & 19  & 60  & 63 \\
Apr  & --  & -- &-- & 14 &31  & 48 & 19  & 65  & 82 \\
May  & --  & -- &-- & 10 &20  & 39 & 15  & 40  & 63 \\
Jun  & --  & -- &-- & 13 &33  & 54 & 18  & 64  & 83 \\
Jul  & --  & -- &-- & 10 &23  & 30 & 22  & 68  & 92 \\
Aug  & --  & -- &-- & 20 &75  & 72 & 20  & 47  & 48 \\
Sep  & 1   & 1  & 1 & 15 &64  & 58 & 15  & 40  & 52 \\
Oct  & 10  & 29 &13 & 23 &174 & 85 & 18  &192  & 97 \\
Nov  & 11  & 24 &11 & 24 &187 & 85 & -- & --   &--\\
Dec  & 13  & 30 &13 & 27 &220 & 88 & -- & --   &--\\
\hline\hline
\end{tabular}
\end{table}

The $E$ column reflects the efficiency of the clear time use and characterizes the gradual decrease of technical problems. The global efficiency for the whole period is 0.49.

\section{OT measurement data reduction}

\subsection{Features of reduction and its modifications}

The conventional DIMM data processing is quite simple: turbulence intensity $J$ is computed directly using eq.~(\ref{eq:KDIMM}) after longitudinal and transversal variances are converted from pixels to radians. Then $J$ is reduced to zenith and expressed as seeing $\beta$ if necessary.

Meanwhile, there is a number of features which upset this simplicity. The first is a finite exposure which suppresses the high-frequency part of the differential image motion. This effect becomes apparent in strong wind in the boundary layer, where the main part of turbulence is usually located, and at high altitudes with wind streams of 25 -- 40\mps. It is common to use a double exposure method when frames of single and double exposures are interlaced and a certain extrapolation \citep{T2002} is exploited to get a `zero-exposure' variance estimation based on theoretical considerations \citep{Martin1987, Soules1996}.

The use of high-speed cameras does not allow for this technique since exposure cannot be changed without stopping the video stream. Instead, the {\sc dimm} program calculates the covariance coefficient between the image coordinates in adjacent frames. Such an approach is theoretically justified by V.\,Kornilov \& B.\,Safonov\footnote{http://curl.sai.msu.ru/mass/download/doc/dimm\_specs.pdf}. Since exposure is as low as 4\,ms the correction does not exceed 0.05 for 90~percent of measurements.
%\bibitem[\protect\citeauthoryear{Kornilov \& Safonov}{2010}]{dm2010} Kornilov V., Safonov B., 2010 http://curl.sai.msu.ru/\\mass/download/doc/dimm\_specs.pdf

A complementary feature is related to the low-frequency domain of the image motion power spectrum. In order to make possible the joint analysis of MASS and DIMM data aimed to restore the full OT profile, the DIMM measurement {\sl basetime} has to be 1 or 2\,s. In this case, part of the power produced by slowly drifting air in the ground layer escapes detection in the DIMM channel.
 
The low-frequency power is estimated using the variance of basetime-derived images separations computed over the {\sl accumtime} span (covering normally 60\,s) and this power is added to the basetime-derived results. This allows to find a trade-off between statistically reliable accumulation time and process stationarity period and to ensure that the method detects a large fraction ($>0.99$) of the motion power.

The third circumstance is that equation~(\ref{eq:KDIMM}) is valid only in the near-field approximation i.e. when $D^2 \gg z \lambda$. This condition is not fulfilled in practice and thus not all the high-altitude turbulence is measured due to wave propagation. This effect was evaluated \citep{T2002,TK07}, but was ignored in practice because its correction implies the knowledge of OT distribution.

The MASS data processing is based on the solution of an equation set of the kind~(\ref{eq:sc}) for a fixed altitude grid under restriction that the solution must be non-negative. This restriction regularises
the solution. The technique of the OT vertical profile restoration from scintillation indices and its features were described by \cite{MD2007} and TKSV.

Experience gained in previous MASS data processing allowed us to significantly modify the algorithms and structure of the data processing utility {\sc atmos} which initially was part of the {\sc Turbina} program. The non-linear minimization method was replaced by a much faster and mathematically proven non-negative least squares method (NNLS). Also, the OT profile restoration by accumtime-averaged scintillation indices was replaced by averaging the OT profiles restored from the 1-s data. Such an algorithm modification ({\sc  atmos-2.96.0a}) reduces the profiles errors and leads to more physically justified turbulence characteristics \citep{exp2010}. Additionally, this permits restoration of an OT profile using denser altitude grid for more accurate localization of the turbulence features. Quality of the profile restoration is controlled by the total residual $R^2$ which is the accumtime-averaged sum of squared residuals of indices.

A further modification ({\sc atmos-2.96.7}) of an algorithm involves the DIMM data in the OT profile restoration. A polychromatic representation of the DIMM weighting function $W_{l,t}(z)$ was deduced by V.\,Kornilov \& B.\,Safonov\footnote{http://curl.sai.msu.ru/mass/download/doc/dimm\_specs.pdf}
which, analogously to eq.~(\ref{eq:sc}), allowed to relate the differential motion variance $\sigma^2$ to the refractive index structural coefficient distribution:
\begin{equation}
\sigma^2_{l,t} = \int_0^\infty C_n^2(z) W_{l,t}(z)\,dz,
\label{eq:dicn}
\end{equation}
The new linear equations system is obtained from the original one for scintillation indices by addition of two equations for longitudinal and transversal components of differential motion and by adding a zero-height point to the altitude grid.

Apart from direct determination of the ground turbulence intensity it is possible now to correct the propagation effect in the differential motion. Non-negativity of the ground layer intensity also stabilizes
the solution.  A detailed description of the new algorithm and its features will follow.

It is important to note that DIMM weighting functions are calculated under assumption that our DIMM measures differential motion which is better matched by aperture wavefront slope in the Zernike polynomial approximation, namely the so called {\sl z-tilt}, rather than gravity centre displacements ({\sl g--tilt}), according to \cite{T2002}. This decreases slightly (by 3 and 5 percent for longitudinal and transversal components, respectively) the derived intensities as compared to those from eqs.~(\ref{eq:KDIMM}) and (\ref{eq:K}).

\begin{figure*}
\centering
\begin{tabular}{cc}
\psfig{figure=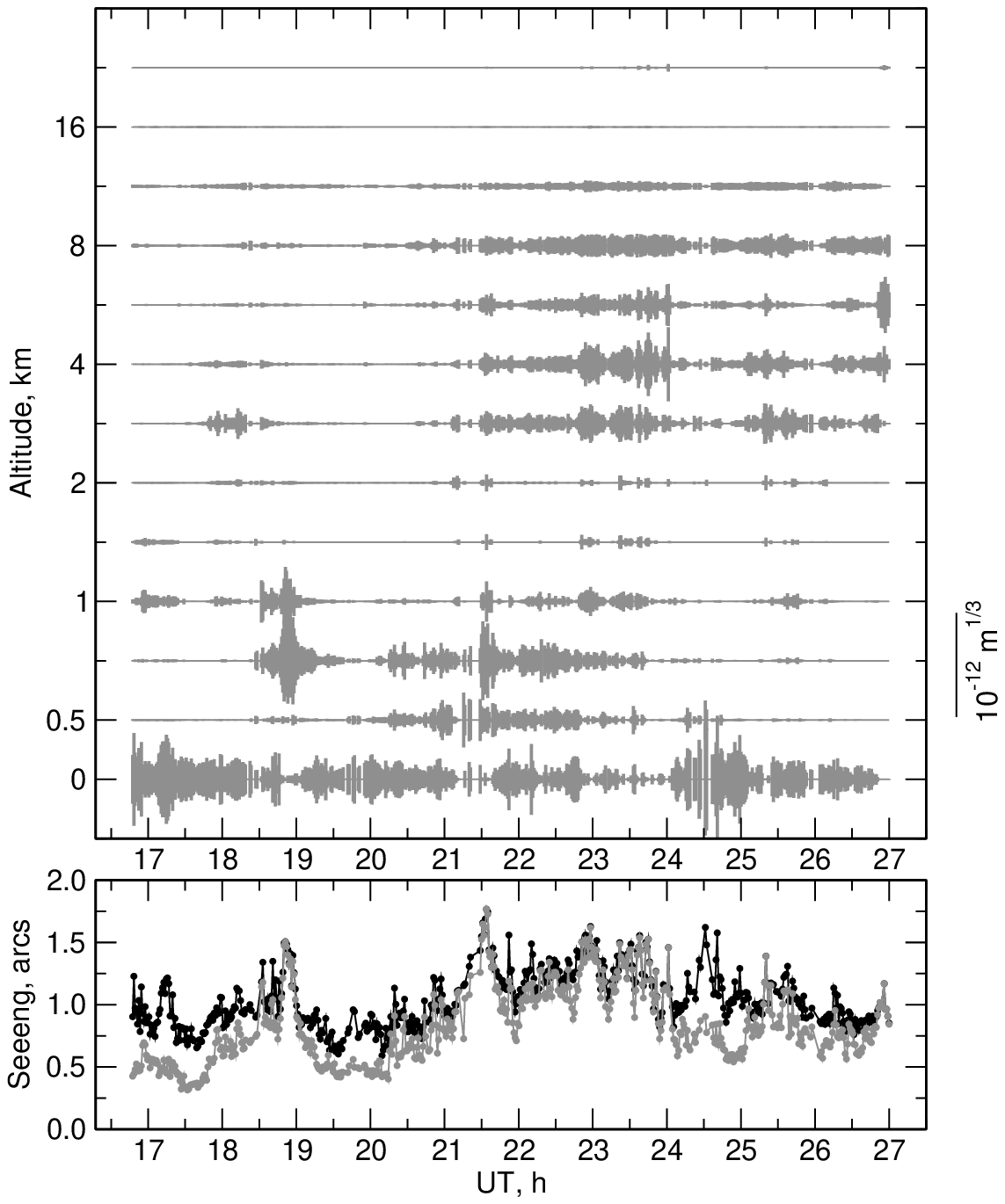,height=9.7cm}&
\psfig{figure=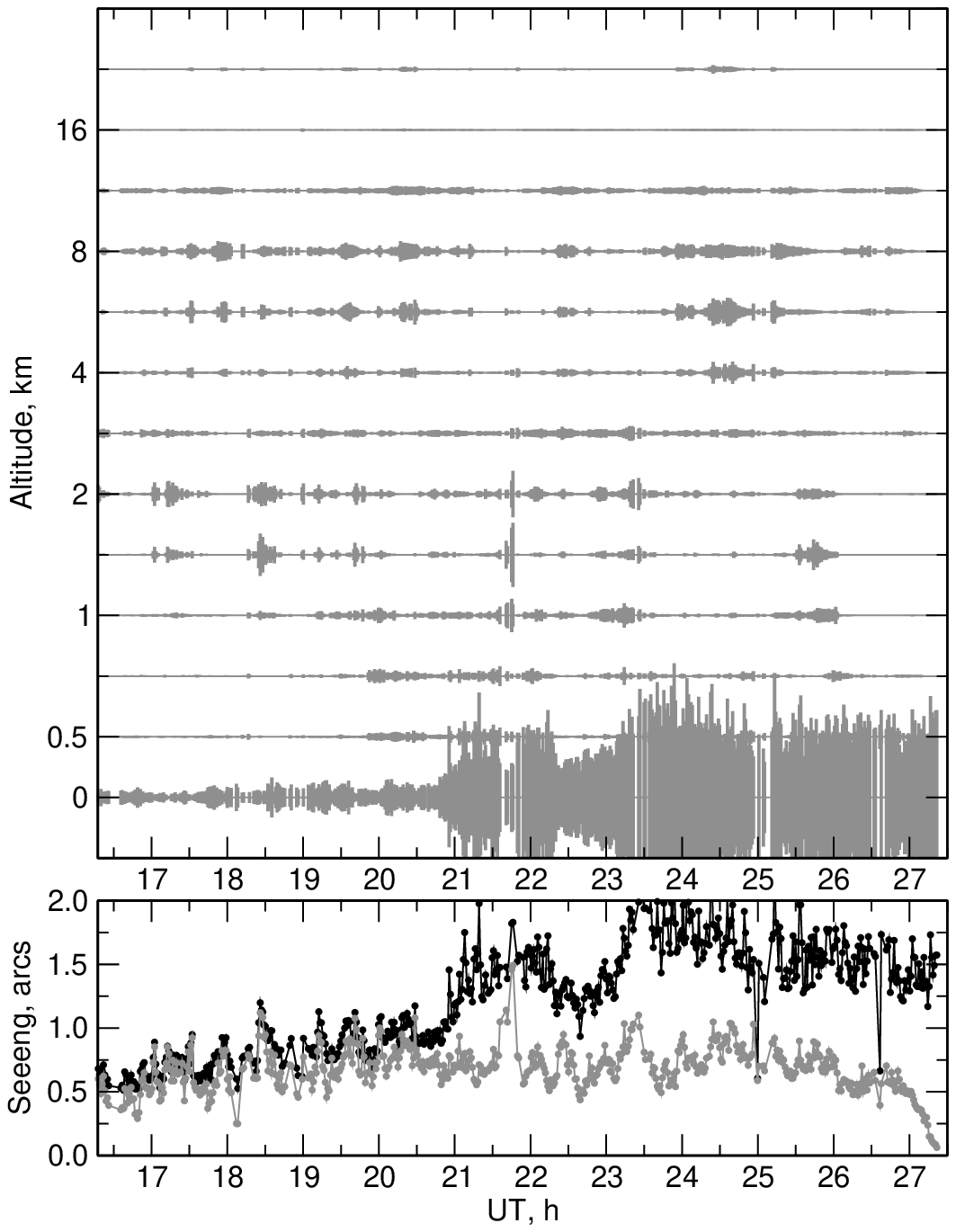,height=9.7cm}\\
\end{tabular}
\caption{ Examples of nightly data processing for 2009 February 19 (left) and 2009 October 24 (right). For each layer, the vertical bars length is proportional to the restored intensity. Scale is shown between the plots.  Lower panels -- the records of seeing $\beta_0$ (whale atmosphere, black points) and of the free atmosphere seeing $\beta_{free}$ (grey points).
\label{fig:images}}
\end{figure*}

\subsection{Data filtering}

The raw MASS data set contains 89189 one-minute star measurements and 6544 background measurements (in 404 nights). When there are DIMM data for the same time intervals they are included in the processing. For some periods they are unavailable, mostly in cases of excessive telescope vibration when a star escapes frequently from the ROI. In this situation no ground layer restoration is performed.

Since the background level is involved in scintillation index computation, it is essential to detect and reject cases when some star was present in the field aperture while the background was recorded. The program accepts the background measurement only if its variance is consistent with solely the photon noise. The target measurements with intensities only marginally exceeding the background level are also rejected.

The data which have passed filtering (88664 meas.) are used in OT profile restoration and integral atmospheric parameters calculation. The residuals analysis is performed to clean out the outlying cases in 1-s OT data set which may affect output statistics.

A further filtration was made while analysing the output results. The events of clouds, the corrector plate fogging, star escapes from the  diaphragm, etc. were detected by the D-channel count drop below 100 pulses per ms and by relative flux error exceeding 0.05. The latter value is much more than the median 0.002 of the relative errors due to scintillation, so the derived statistics were not biased. These checks have rejected 0.9~percent of data. Finally, extinction star measurements were thrown away if performed at air mass higher than 1.38. MASS instrumental problems in the period between July 19 and 23 (2008) were detected by abnormal C/D flux ratios.

The residuals analysis has demonstrated that including DIMM data in the profile restoration significantly increases the total residual $R^2$ (from 1.46 to 2.62 for the global median) in spite of the weak coupling of MASS and DIMM data. The reason was an inconsistency between longitudinal and transversal image-moion variance. When only one DIMM equation is included in the system (either $l$ or $t$), the $R^2$ drops back to almost its original, MASS-only level.

The distribution of residuals is similar to the one without DIMM data but starting at $R^2\approx 4$ it shows a prominent flat tail. In these cases the actual data come into severe conflict with the model. Their inspection allows to propose two main reasons: 1) sometimes MASS determines stronger turbulence than follows from the DIMM data, and 2) longitudinal and transverse components of the differential motion are mutually incompatible.

The first situation is discussed in a number of papers as `overshoots' or `negative ground layer' problem and is considered in Section~\ref{sec:struct}. Although we made some progress in treatment of those cases (see above) and they are encountered now quite rarely, they have not disappeared completely. The second reason became especially prominent in June 2009 due to optics misalignment when the median residual grew up to $\approx 5$.

To filter out the completely inadequate solutions, we adopted a threshold residual $R^2=12.4$ which corresponds to 0.5~percent cases which appear to be indeed marginal. Finally, a total of 85512 OT measurements have passed all filters and are included in the statistics discussed below.

\section{Turbulence measurement results}

\subsection{Seeing statistics}
%corr: Page 11, last line of the first paragraph of section 8.1 remove "its" in front of "direction"
Vertical OT distributions were restored on the logarithmic altitude grid of 13 layers at 0, 0.5, 0.71, 1, 1.41, 2, 2.82, 4, 5.66, 8, 11.3, 16 and 22.6~km above the observatory. This grid is based on the conventional MASS grid (see TKSV) with intermediate and zero-height nodes inserted. As an example, in Fig.~\ref{fig:images} we present processing results for two nights which show different OT characters: on February, 19 (2009) the conditions got worse after midnight when turbulence intensified at 3 -- 8\,km, and on October, 24 it was a ground layer which was responsible for the seeing deterioration. The wind data for this night report that at 22\,h the ground wind became stronger and changed direction.

As previously said, the OT power can be characterised by different measures: intensity $J$, seeing $\beta$, or by the Fried radius $r_0$. The processing results are expressed in terms of $J$ but further on we will also use seeing, depending on the context. Quantities related to the whole atmosphere will be referred to as $J_{total}$ and $\beta_0$.

The discussion of OT at a given site and the subsequent use of results involve normally the a distinction between ground and boundary layers. Here we consider layers at 0.5\,km and above as free atmosphere. The respective turbulence intensity is denoted $J_{free}$.

\begin{figure}
\centering
\psfig{figure=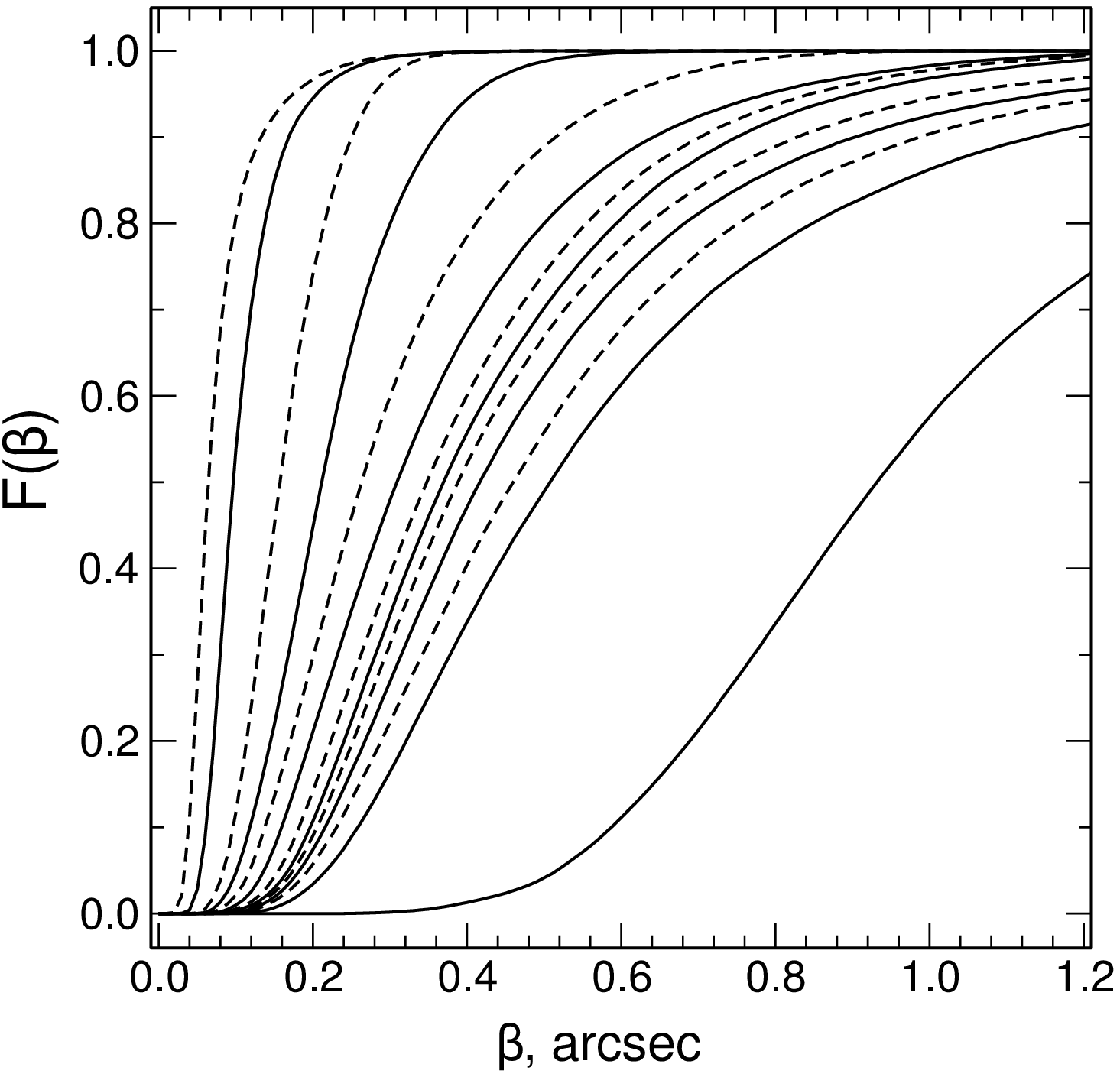,height=6.5cm}

\bigskip

\psfig{figure=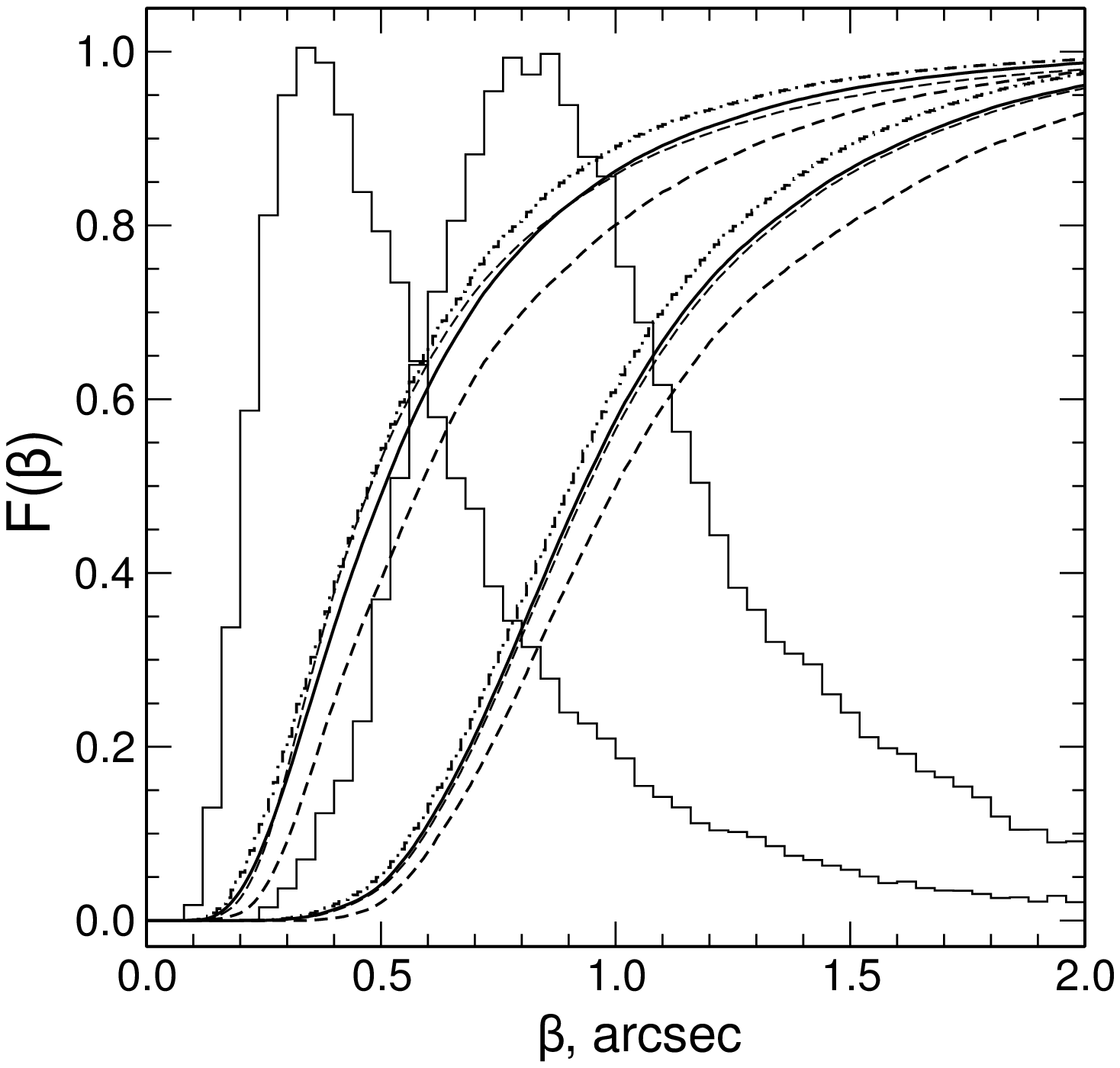,height=6.5cm}
\caption{ Top: integral distributions of the total OT intensity  expressed as seeing $\beta_h$ computed for atmosphere layers at $h$ and above. Solid lines from right to left: total atmosphere, 0.5\,km  and higher, 1\,km and higher, etc. for 2, 4, 8 and 16\,km levels.
Dashed lines are the same for intermediate heights. Bottom: integral  distributions of the total image quality $\beta_0$ (right solid line) for the total data set, for winter (dash-dotted) and summer (dashed) seasons. Thin dashed curve shows the distribution of $\beta_0$ computed by the conventional formula. Leftmost solid and thin curves present $\beta_{free}$ derived from OT profiles and from the zero-order moment (integral atmosphere parameter $M_0$). Thin step-like curves are differential $\beta_0$ and $\beta_{free}$  distributions.
\label{fig:distribs}}
\end{figure}

%corr: Page 11, second to last full paragraph "height" instead of "heihgt"
A convenient and common way to present statistics of the OT vertical distribution is a cumulative distribution for the sum OT intensities from a given height $h$ to the upper boundary of the atmosphere. In other words these curves tell us what is the image quality for an observer at a given $h$. As shown in \citep{exp2010}, these integrals possess a higher precision than intensities in individual layers since errors in adjacent layers anti-correlate.

Such cumulative distributions expressed in terms of seeing are shown in Fig.~\ref{fig:distribs} (top). Their characteristic points are listed in Table~\ref{tab:seeing}. The graph at the bottom contains most relevant distributions $\beta_0$ and $\beta_{free}$. The distribution of $\beta_0$ derived from the OT profiles coincides with the one computed by the conventional DIMM formula to within $\pm0.01$~arcsec. For the free atmosphere, the difference between the OT profile and the integral moment-derived $\beta_{free}$ is more visible but the median difference is 0.04~arcsec only.

The same plot shows the `winter' (September -- February) and `summer' (March -- August) seasonal distributions. Since the winter data are 2.4 times more voluminous, the `winter' curve tends to fit the global one, differing by $0.02$~arcsec only. The `summer' seeing has a median of 1.00~arcsec.

Differential $\beta_0$ and $\beta_{free}$ distributions allow to determine the {\sl mode} which characterizes the most probable value of seeing at Shatdzhatmaz. Since the curves are positively asymmetric
the modes are considerably less than the medians and equal to 0.82 and 0.35~arcsec, respectively.

\begin{table}
\caption{The quartiles of the OT intensity $J$ (Q2 is a median) which is generated by the given altitude layer (in km) and the layers above and of the respective seeing $\beta$. \label{tab:seeing}}
\centering
\begin{tabular}{l|rrr|rrr}
\hline\hline
 &\multicolumn{3}{c|}{$\beta^{\prime\prime}$} & \multicolumn{3}{c}{$J, \times10^{-14}\mbox{ m}^{1/3}$} \\
Layer   & Q1 & Q2 & Q3 & Q1 & Q2 & Q3 \\
\hline
Whole    &  0.731 & 0.932 & 1.222 & 40.3 & 60.4 & 94.8\\
0.5+      &  0.349 & 0.508 & 0.760 & 11.8 & 21.9 & 43.0\\
0.71+     &  0.315 & 0.460 & 0.680 & 9.89 & 18.6 & 35.7\\
1.0+      &  0.291 & 0.417 & 0.617 & 8.68 & 17.0 & 30.4\\
1.41+     &  0.273 & 0.388 & 0.575 & 7.82 & 15.8 & 27.0\\
2.0+      &  0.261 & 0.370 & 0.542 & 7.23 & 14.0 & 24.5\\
2.82+     &  0.243 & 0.348 & 0.507 & 6.41 & 12.9 & 21.8\\
4.0+      &  0.214 & 0.309 & 0.455 & 5.20 & 9.58 & 18.2\\
5.66+     &  0.186 & 0.263 & 0.375 & 4.12 & 7.33 & 13.3\\
8.0+      &  0.157 & 0.212 & 0.279 & 3.10 & 5.10 & 8.10\\
11.3+     &  0.121 & 0.157 & 0.202 & 2.02 & 3.11 & 4.72\\
16+       &  0.076 & 0.097 & 0.128 & 0.92 & 1.38 & 2.19\\
22.6     &  0.049 & 0.064 & 0.090 & 0.45 & 0.70 & 1.22\\
\hline\hline
\end{tabular}
\end{table}

It should be stressed that integral distributions as well as the dependence of mean or median layer intensity on its altitude are useful for general statistical OT evaluation at a given site. Meanwhile they could hardly help to understand the atmospheric processes since they do not reflect correlations between turbulent layers.

Of a considerable interest are the seasonal variations of characteristic OT quantities. Although we cover only two years of observations at Shatdzhatmaz and the monthly behaviour isn't well defined, we may already confidently conclude from Table~\ref{tab:seasons} that October and November deliver the best conditions both in total turbulence and in the free atmosphere. On the other hand, the end of winter (March) shows a remarkably strong turbulence.

\begin{table}
\caption{Monthly seeing distributions for the whole atmosphere and for layers at 1\,km and above (in arcseconds). $N$ is the number of measurements in a given month.
\label{tab:seasons}}
\centering
\begin{tabular}{l|rrrrrrr}
\hline\hline
       &  & \multicolumn{3}{c}{Total seeing $\beta_0$} & \multicolumn{3}{c}{Free seeing $\beta_1$} \\
Months    &N, $10^3$& Q1 & Q2 & Q3 & Q1 & Q2 & Q3 \\
\hline
Jan       &  5.0  & 0.77 & 0.98 & 1.28 & 0.27 & 0.42 & 0.63 \\
Feb       &  7.8  & 0.82 & 1.00 & 1.28 & 0.37 & 0.56 & 0.87 \\
Mar       &  3.9  & 1.30 & 1.70 & 2.06 & 0.55 & 0.88 & 1.35 \\
Apr       &  4.7  & 0.75 & 1.02 & 1.45 & 0.28 & 0.39 & 0.58 \\
May       &  2.9  & 0.73 & 0.95 & 1.38 & 0.30 & 0.39 & 0.60 \\
Jun       &  4.3  & 0.78 & 0.91 & 1.10 & 0.35 & 0.43 & 0.54 \\
Jul       &  3.6  & 0.80 & 0.93 & 1.12 & 0.38 & 0.53 & 0.71 \\
Aug       &  5.6  & 0.70 & 0.95 & 1.16 & 0.31 & 0.46 & 0.65 \\
Sep       &  5.3  & 0.80 & 0.97 & 1.21 & 0.29 & 0.38 & 0.55 \\
Oct       &  18.5 & 0.64 & 0.82 & 1.04 & 0.23 & 0.34 & 0.50 \\
Nov       &  11.0 & 0.65 & 0.84 & 1.16 & 0.26 & 0.36 & 0.52 \\
Dec       &  13.0 & 0.78 & 0.97 & 1.25 & 0.32 & 0.44 & 0.61 \\
\hline\hline
\end{tabular}
\end{table}

\subsection{Ground layer properties}
\label{sec:gl}

Particularly important is the behaviour and characteristics of the ground layer as a dominant contributor of turbulent atmosphere. In principle, simultaneous MASS and DIMM measurements allow to extract its power as a difference of $J_{total}$ and $J_{free}$ values which in fact originate from different instruments and methods. It's evident that systematic errors of their determination will amplify in a difference $J_{total} - J_{free}$. In cases when the ground layer dominates, this is not that critical, but when it is weak or when the high-altitude turbulence is strong the problems could arise.

Potential systematic biases of DIMM results are already described and taken into account in the data processing. Meanwhile, the noticed uncertainty in the slope type which is sensed by the particular DIMM
instrument is not still experimentally examined. It is probably neither z-tilt nor g-tilt but some intermediate characteristic. The maximum resulting uncertainty is about 17~percent in intensity or 10~percent in seeing.

There are biases in the MASS data processing as well, described in detail by \cite{TK07}. The most significant is a deviation from the weak-perturbation approximation in the case of strong high turbulence,
causing near-saturated scintillation to develop in small apertures. Although authors do provide some empirical correction, the derived solution is formally worse by the $R^2$ criterium. That's why the {\sc atmos-2.96} data processor makes only the following corrections of scintillation indices:
\begin{equation}
s^2 = \ln(\tilde s^2+1),
\label{eq:log}
\end{equation}
where $\tilde s^2$ is the computed index in a sense $\left< \delta I^2\right>/\left<I\right>^2$. The reason is that the MASS weighting functions are defined for the $\chi = \ln A$ fluctuations where $A$ is a light wave amplitude and the scintillation index is defined as $s^2 = 4\sigma^2$, where $\sigma^2$ is a variance of $\chi$. The measured light intensity has a log-normal distribution and for a strong scintillation ($s^2\sim 0.5$) this leads to a 20~percent error.

\begin{figure}
\centering
%\begin{tabular}{cc}
\psfig{figure=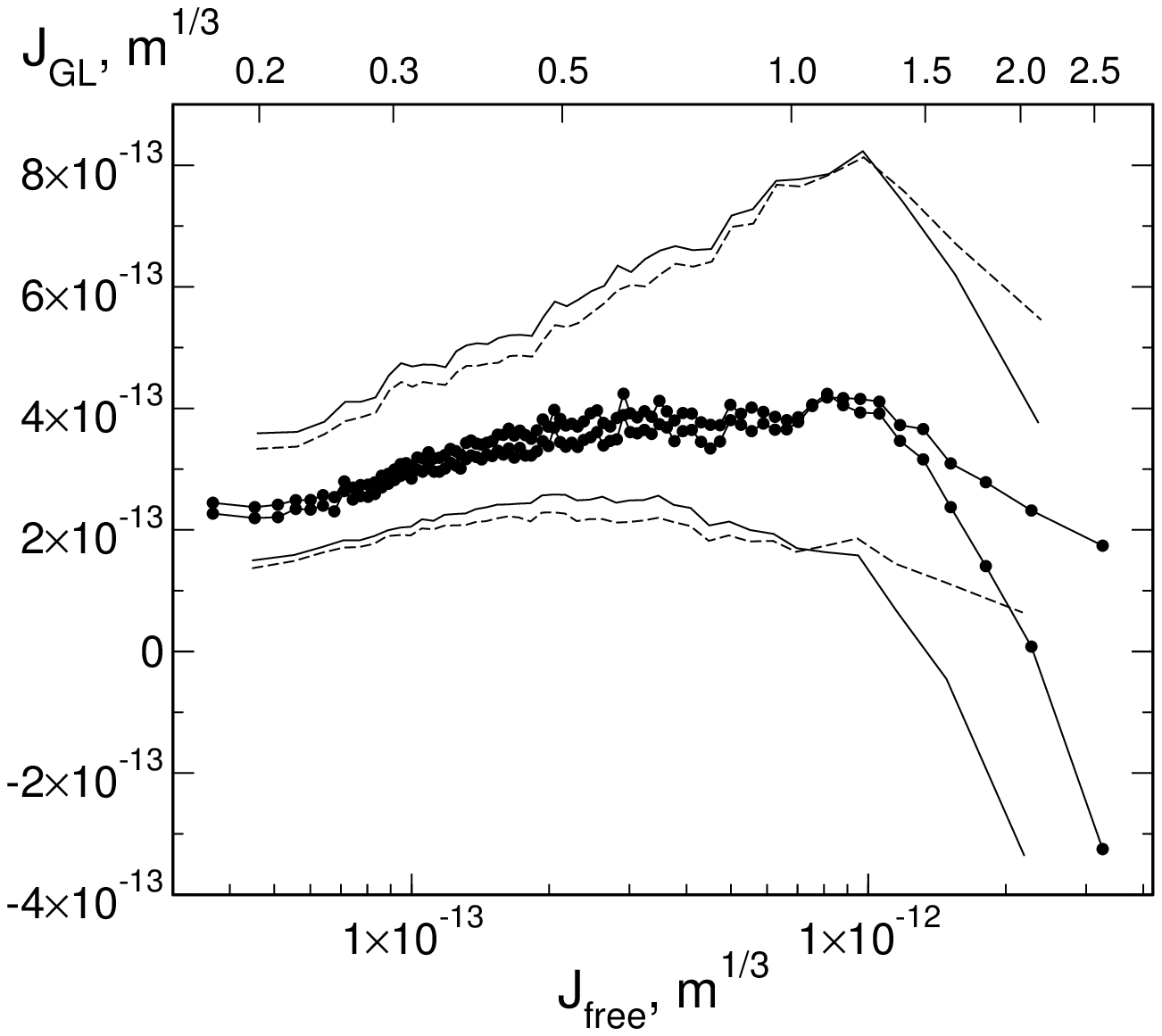,width=7.4cm}

\bigskip

\hspace*{0.4cm}\psfig{figure=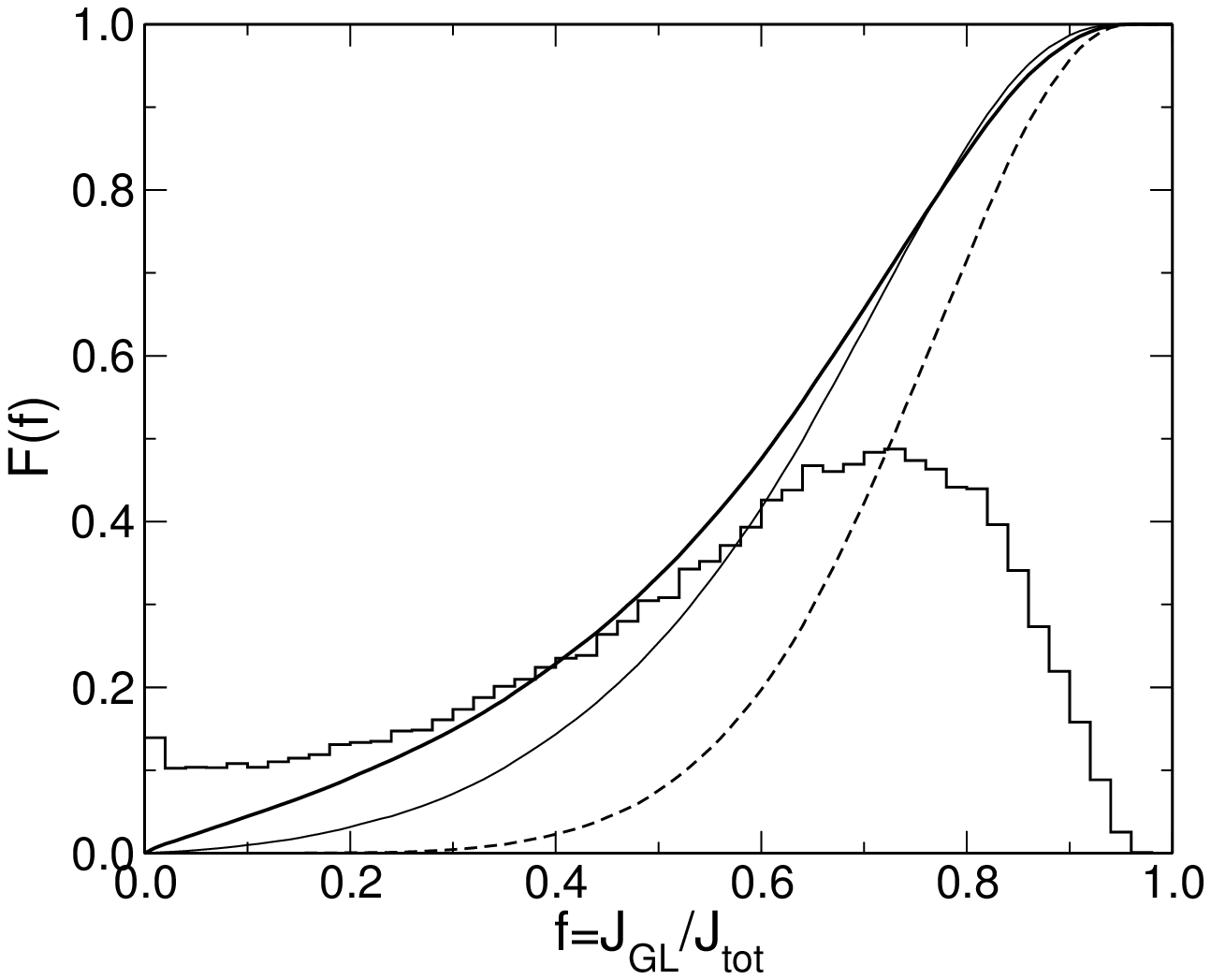,width=7.4cm}
%\end{tabular}
\caption{Top: dependence of a median ground layer power as obtained from integrals $J_{total}$ and $J_{free}$ (black dots) and from OT profiles (open circles) on the intensity $J_{free}$. Thin solid and dashed curves depict first and third quartiles of respective distributions. Bottom: integral distribution of the ground layer intensity fraction for the total data set (thick solid line), the $\beta_0 < 0.93$~arcsec subset (thin solid) and the $\beta_{free} < 0.50$~arcsec subset (dashed). The stair curve is differential total distributions.
\label{fig:gl}}
\end{figure}

This problem was also discussed in the past but consideration of a nearly saturated scintillation refrained us from additional diminishing of scintillation indices. Nevertheless, the real data examination has shown that the described correction works in the right way: the OT power becomes weaker and the turbulence itself tends to localize higher.

The first estimation of the ground layer intensity $\hat J_{GL}$ was obtained from the altitude turbulence moments and the conventional formula~(\ref{eq:KDIMM}). Our study reveals that application of the correction~(\ref{eq:log}) decreases the fraction of problematic data from $\approx0.10$ to $< 0.04$ and such data do not originate from random errors of the ground layer intensity determination.

In Fig.~\ref{fig:gl} we show the dependence of the median values $\hat J_{GL}$ (calculated over groups of 1000 points) on $J_{free}$. It is clear that in conditions of a very weak (about $4\cdot 10^{-14}\mbox{ m}^{1/3}$) free atmosphere OT $\hat J_{GL}\approx 2\cdot 10^{-13}\mbox{ m}^{1/3}$ (nearly 0.5~arcsec), i.e. it exceeds significantly the $J_{free}$ intensity. As turbulence gets stronger, $\hat J_{GL}$ increases. Such a relation indicates a weak and most probably indirect interrelation between GL and free atmosphere. Further, the ground layer intensity stabilizes at the level $4\cdot 10^{-13}\mbox{ m}^{1/3}$ ($\approx 0.7$~arcsec) and remains unchanged while $J_{free}<1\cdot 10^{-12}\mbox{ m}^{1/3}$ ($\approx
1.3$~arcsec).

The ground layer intensity $\check J_{GL}$ derived from the OT profile restoration behaves in a similar way. Meanwhile, when the free atmosphere turbulence becomes too strong and the above-mentioned systematical effects cause the $\hat J_{GL}$ to drop tending to negative values, $\check J_{GL}$ remains positive.

Such a difference is explained by the propagation effects becoming significant for DIMM measurements in dominant free atmosphere turbulence conditions. An additional circumstance is a non-negativity restrictions applied in OT restoration which act similarly to saturated indices correction, i.e. decreasing and rising the resulting turbulence.

In Fig.~\ref{fig:gl} (bottom) we plot the differential and integral distributions of the fractional power $f$ of the ground layer. The integral distribution of all measurements demonstrates the systematics in the small $f$ domain, with a small excess of zero values. This shows that the implemented correction is not sufficient, although it prevents negative $\check J_{GL}$ values.

These effects are diminish in the $f$ distribution for the cases of $\beta_0$ better than median and are absent in the $\beta_{free} < 0.50$~arcsec conditions. Note that the distribution is highly asymmetric due to the nature of $f$ being a ratio, so the mode and the median are largely different. For the total data set, the mode and the median are 0.71 and 0.61, for the sample $\beta_0 < 0.93$~arcsec they are 0.71 and 0.64, and when $\beta_{free} < 0.50$~arcsec they are 0.78 and 0.73.

One can expect that a complete correction for saturation would make the total data distribution closer to the good-images case, i.e. the median fraction of the ground layer will increase to $\approx 0.65$. This value is typical for many summits (at Las Campanas observatory $f=0.63$ as \cite{gmt2008} quote) except for those good sites which have a very thin ground layer, e.g. $f=0.46$ at Mauna Kea \citep{tmtV}.

\begin{figure}
\centering
%\begin{tabular}{cc}
\psfig{figure=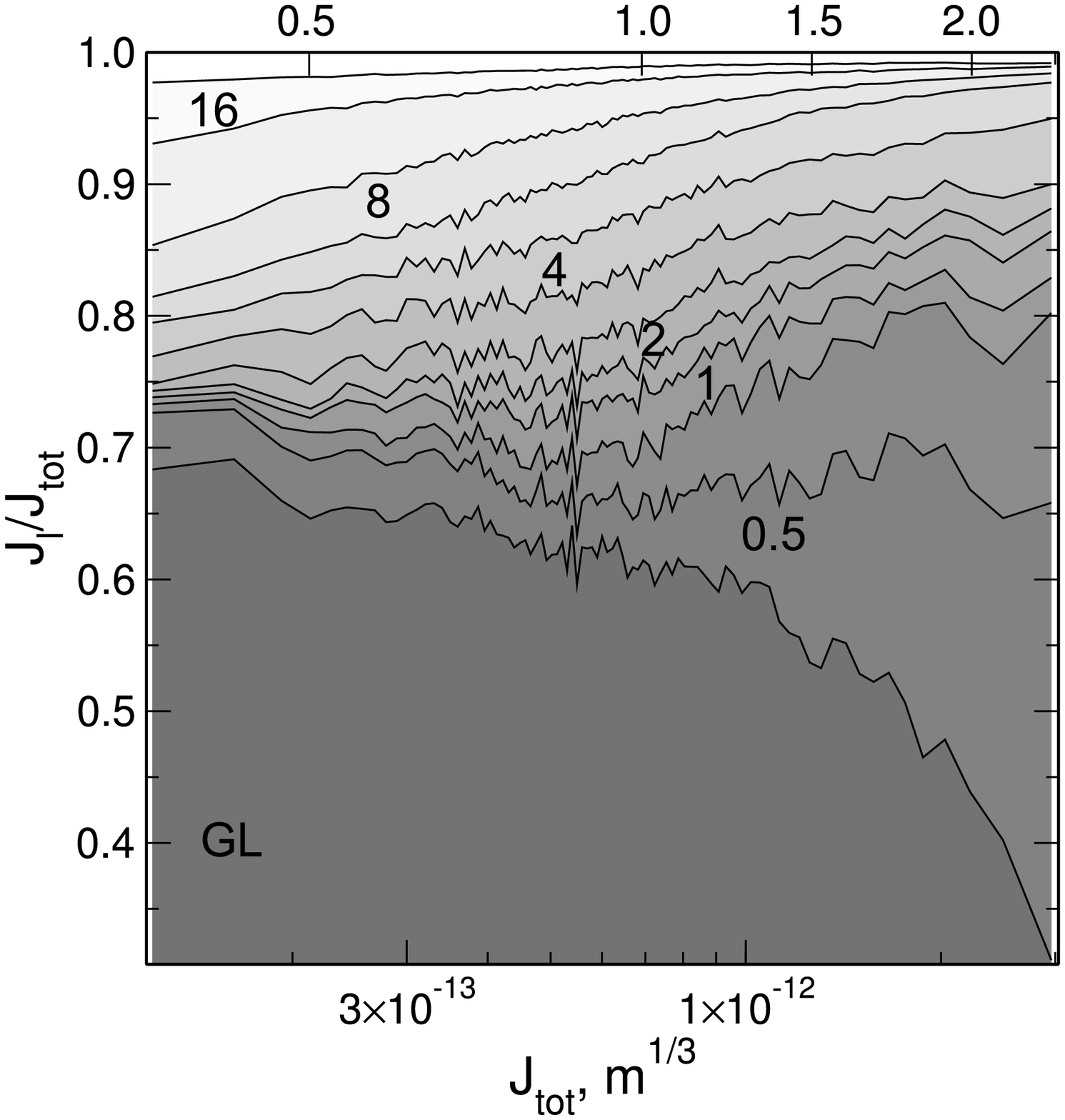,height=7cm}

\medskip

\psfig{figure=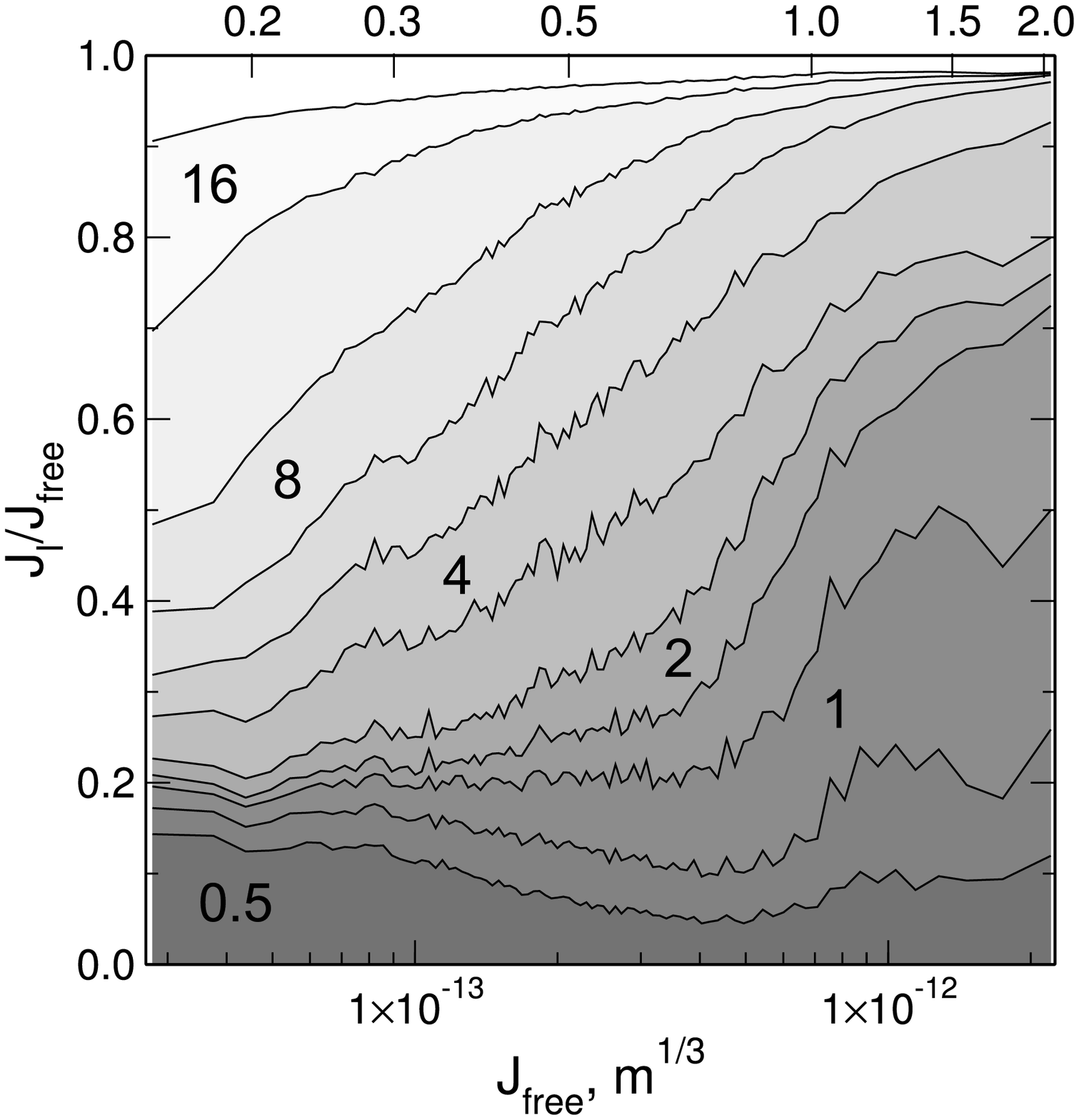,height=7cm}
%\end{tabular}
\caption{Structure of turbulent atmosphere as it depends on the total intensity $J$. The stripe thickness reflects its fraction of the total. Top: the whole atmosphere structure as a function of $J_{total}$ (i.e. including GL). Bottom: the free atmosphere structure as a function of $J_{free}$. Alternative (upper) X-axes provide the seeing scale in arcseconds.
\label{fig:fracs}}
\end{figure}

\subsection{Structure of turbulent atmosphere}
\label{sec:struct}

Many authors \cite[e.g. see][]{CP2006} notice that the OT distribution character depends on the total intensity. In order to study the properties of the OT structure on the basis of the restored profiles, the median layer intensity fractions were plotted in Fig.~\ref{fig:fracs} for the whole atmosphere and for the free one. Note that in these pictures a layer fraction is the {\sl width} of the respective stripe.

In the top graph the full atmosphere structure is plotted against $J_{total}$. The abrupt lowering of the GL fraction above $J_{total} = 1\cdot 10^{-12}\mbox{ m}^{1/3}$ is related to systematic effects in its determination as discussed in the previous section. In the weaker turbulence domain the real diminishing of its fraction from 0.7 to 0.6 is seen. The sum intensity of the ground and 0.5-km layers accounts for about 0.70 of the total power in the whole $J_{total}$ range.

It is informative that strengthening of the total turbulence is powered by the growth of lower 0.5 and 0.7\,km layers which belong to the boundary region of atmosphere and also by the 4 and 6\,km layers. This is accompanied by diminishing of the 8 to 16\,km layers contribution which normally dominate when the total OT is weak. The increasing role of the lower layers may be ascribed to the thickening of the ground layer in bad conditions.

Changes in the free atmosphere structure as they evolve with its overall power are shown in the bottom plot of Fig.~\ref{fig:fracs}. The general tendency is an increase of the fraction of low layers with increasing total intencity o fthe OT in the free atmosphere. The whole range can be sub-divided in three domains:
$\beta_{free}<0.3$~arcsec, $\beta_{free}>1.0$~arcsec and an intermediate. The leftmost region corresponding to the best conditions shows the layers above 8\,km dominating, in the intermediate conditions the major OT contribution descends to 2 -- 8\,km, and in the worst situation -- down to 0.7 -- 1.4\,km.

\subsection{Integral parameters}

In addition to the OT profile restoration, the computed scintillation indices also serve for calculation of the altitude moments of turbulence as described in TKSV:
\begin{equation}
M_{\alpha} = \int C_n^2(h)h^\alpha\, dh
\label{eq:Mdef}
\end{equation}
for $\alpha = 0, 1, 5/3, 2$. These moments are used for integral atmospheric parameters determination: free atmosphere seeing $\beta_{free}^{5/3} \sim M_0$, effective OT altitude $h_{eff} \sim M_1$, isoplanatic angle $\theta_0^{-5/3} \sim M_{5/3}$ and isokinetic angle $\theta_k^{-2} \sim M_2$. From the $s^2_A$ scintillation indices obtained with a tripled exposure the atmospheric coherence time $\tau_0$ is estimated as proposed by \cite{T2002time}. In these terms, the seeing $\beta_0$ obtained by the formula (\ref{eq:KDIMM}) is also a kind of integral parameter.

\begin{figure}
\centering
\psfig{figure=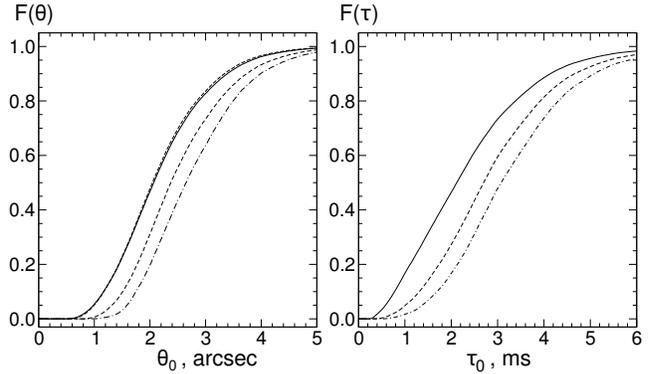,width=8.4cm}
\caption{ Left: integral isoplanatic angle $\theta_0$ distributions for the whole data set (solid line), a $\beta_0 < \mbox{median}$ subset (dashed) and $\beta_0 < \mbox{25\%}$ (dashed-dotted). The leftmost dashed line depicts $\theta_0$ by OT profiles. Right: integral distributions of atmospheric constant $\tau_0$ for the same data samples.
\label{fig:integ1}}
\end{figure}

Atmospheric moments are computed as linear scintillation indices combinations where the related coefficients are derived from the needed $h^\alpha$ approximations by the weighting functions $Q(h)$
(TKSV). The feasible precision ($\delta < 0.05$) of this decomposition is attained at altitudes $h > 0.4$\,km which is sufficient because lower layers produce almost no scintillation.

The isoplanatic angle determined from the $M_{5/3}$ agrees well with the angles computed from OT profiles. The regression slope is 0.98 with a rms scatter of 0.03~arcsec around it. The median $\theta_0$ from profiles is 2.03~arcsec, differing by less than 2 percent from the $\theta_0$ median calculated from the moments. This fact adds confidence to the results of OT restoration by the implemented algorithm.

\begin{table}
\caption{Cumulative statistics of integral atmospheric parameters. Isoplanatic and isokinetic angles $\theta_0$ and $\theta_k$ are in arcseconds.
\label{tab:iso}}
\centering
\begin{tabular}{l|rrr|rrr}
\hline\hline
  &\multicolumn{3}{c}{$\theta_0$} & \multicolumn{3}{c}{$\theta_k$} \\
                                    & Q1 & Q2 & Q3  & Q1 & Q2 & Q3 \\
\hline
whole set              & 1.56  & 2.07  &  2.70  &  5.54 & 7.00  & 8.67  \\
best 50\% of $\beta_0$ & 1.88  & 2.38  &  3.06  &  6.42 & 7.78  & 9.52  \\
best 25\% of $\beta_0$ & 2.11  & 2.65  &  3.33  &  7.04 & 8.46  & 10.17 \\
\hline\hline
\end{tabular}
\end{table}

\begin{table}
\caption{Cumulative statistics of the time constant  $\tau_0$, in ms.
\label{tab:tau}}
\centering
\begin{tabular}{l|rrr|rrr}
\hline\hline
   &\multicolumn{3}{|c|}{$\tau_0$} & \multicolumn{3}{c}{$\tau_0$ correct.} \\
                              & Q1 & Q2 & Q3   & Q1 & Q2 & Q3 \\
\hline
whole set               & 1.28  & 2.12  &  3.09 &  1.59 & 2.58  & 3.70  \\
best 50\% of $\beta_0$  & 1.92  & 2.70  &  3.67 &  2.40 & 3.34  & 4.46  \\
best 25\% of $\beta_0$  & 2.32  & 3.08  &  4.06 &  --   & --   & --   \\
\hline\hline
\end{tabular}
\end{table}

Integral $\theta_0$ distributions for three samples: the full scope of data, the cases of $\beta_0 < 0.93$~arcsec and of $\beta_0 < 0.73$~arcsec are shown in Fig.~\ref{fig:integ1} (left) and their characteristics are listed in Table~\ref{tab:iso}. Since $\theta_0$ is essentially driven by turbulence in the upper atmosphere, its value is tied with the altitude of the summit. Given this circumstance, the
obtained value is as good as at many other observatories.

For example, our MASS measurements at Maidanak \citep{maid2005} derived the characteristic $\theta_0$ to be 2.19~arcsec which varied seasonally from 1.92 to 2.35~arcsec. \cite{ctio4} and \cite{CP2006} list $\theta_0 = 1.56$ and 2.22~arcsec for Cerro Tololo and Cerro Pachon, respectively.

Values of the AO time constant $\tau_0$ delivered by MASS and listed in Table~\ref{tab:tau} do not correspond to the standard definition of this quantity, as noticed by many authors. The estimates obtained by this method require a certain correction \citep[see, for  example,][]{tmtVII}. For correction of individual values, we used the formula~(3) from the updated study of A.Tokovinin\footnote{http://www.ctio.noao.edu/\symbol{126}atokovin/profiler/\\\rule{0.5cm}{0pt}/timeconstnew.pdf}. The statistics for corrected $\tau_0$ are also shown in Table~\ref{tab:tau}.

Isokinetic angle $\theta_k$ which provides the angular size of the wavefront slopes correlation has a special property to depend not only on the wavelength but also on the telescope aperture. According to \cite{Hardy} it relates to the turbulence second moment by the following expression:
\begin{equation}
\theta_k^{-2} = 0.688\left(\frac{2\pi}{\lambda}\right)^2 D^{-1/3} M_2
\end{equation}
In Table~\ref{tab:iso} the characteristics of the observed $\theta_k$ distribution are given for the case of $\lambda=500$\,nm and the telescope $D=2.5$\,m. The comparison with the value derived from OT profiles shows a very good coincidence; the median isokinetic angle equals 7.01~arcsec. Explicit measurement of $\theta_k$ is omitted in a majority of site testing campaigns since it strongly correlates with the isoplanatic angle and depends (although weakly) on the aperture size.

\section{Discussion and conclusions}

\subsection{Results discussion}
Detailed analysis and interpretation of the obtained results extend beyond the framework of the current study. Here we limit ourselves to two examples of the potential use of the provided information.

Various statistics of seeing and other relevant parameters are not sufficient for comprehensive characterization of the turbulent atmosphere properties at a given site. Of great importance are the
temporal characteristics describing the typical rate of OT evolution, its time-scales of stability, and so on. These issues will be examined in forthcoming papers. In the current study we provide only a crude
estimate of the temporal stability of seeing in a form suitable for a typical astronomical application.

%REF: Page 15, fig 14. /Delta on the y-axis again shows as a crossed circle.
\begin{figure}
\centering
\psfig{figure=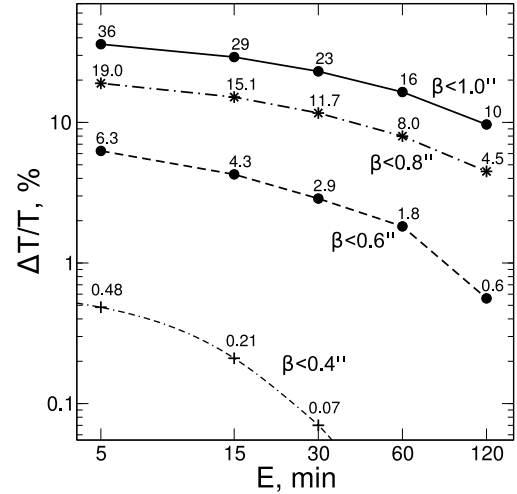,height=6.5cm}
\caption{Fraction of available continuous astronomical exposures of the duration $E$ with a given limit of seeing $\beta_0$.
\label{fig:expos}}
\end{figure}

In Fig.~\ref{fig:expos} we show the fractions of observing time suitable to deliver continuous periods $E$ of the seeing better than a selected threshold. For example, the 30-minutes exposures with an image quality better than 0.6~arcsec are possible in 3 percent of the available observing time, which corresponds to 80 such exposures per year. If we restrict ourselves to 15-minutes exposures, we will get more than 230 frames per year. Such a statistic is evidently over-constrained, since short-term image deteriorations do not spoil the images that seriously, but these `almost good' exposures would not be counted. A similar prognosis of OT stability was presented in the paper of \cite{tmt-temp}.

\begin{figure}
\centering
\psfig{figure=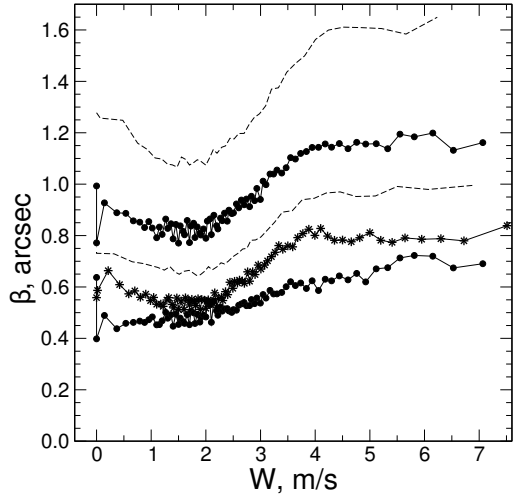,height=6.5cm}\\
\caption{Dependence of the median seeing $\beta_0$ (black dots),  $\beta_{free}$ (open circles) and $\beta_{GL}$ (stars) on the ground  wind speed $W$. Dashed lines depict the first and third quartiles for $\beta_0$.
\label{fig:weffects}}
\end{figure}

Turbulence generation in the ground layer is usually related to mixing of local air masses by the wind. The characteristic feature of this phenomenon is the dependence of turbulence intensity on the ground wind speed. Such a correlation is observed at many sites, as reported by \cite{gmt2008} and \cite{tmtV}. In Fig.~\ref{fig:weffects} we plot the median seeing for the ground layer, free, and full atmosphere as a function of the recorded ground wind speed. These medians were calculated over groups of 1000 points. The effect is quite prominent, as reflected by the median values spread and interquartile range shown for $\beta_0$.

It is also evident from the $\beta_{GL}$ curve that strengthening of the ground wind causes first the decrease of the detected seeing which minimizes at speeds 1.5 -- 2\mps\ and then starts getting worse with
speeds up to 4\mps. A similar effect is noted by \cite{tmtV} at all monitored sites except San Pedro Martir. A further wind speed growth has already no effect on the ground turbulence but apparently increases the boundary layer seeing at 0.5\,km and higher. It is remarkable that the minimum $\beta_{GL}$ corresponds to the most probable wind speed at Shatdzhatmaz in clear night conditions, 2\mps.

\subsection{Conclusions}

The measured basic site characteristics and OT profiles have confirmed the efficiency of the developed and built automatic site monitor. The OT data amount and quality exceed very significantly the results of
our previous 3-year campaign performed at Mt.~Maidanak in 2005 -- 2007 in the manual observations mode \citep{maid2005}. This convincingly justifies the time and resources invested in the complete automation
of astronomical monitoring measurements.

Here we list the most significant site characteristics determined in the site testing program at Mt.~Shatdzhatmaz in 2007 -- 2009:
\begin{itemize}
\item The annual clear astronomical night time is 1343\,hours, the clear nautical night time is 1546\,h, under a rather strict clear  sky criterion.
\item The maximum of the clear sky amount is observed from mid-September to mid-March, where about 70~percent of the clear weather is concentrated.
\item The mean clear night-time temperature is $+1.8^\circ$C, which  implies a moderate water vapour content in the atmosphere.
\item On clear nights, the median precipitable water vapour is 7.75\,mm.
\item The day-to-night temperature contrast is $\approx3.4^\circ$C for clear-sky weather.
\item The median ground wind speed is 2.3\mps\ with a preferred direction from the West.
\item The median seeing $\beta_0$ is 0.93~arcsec. In 25~percent of time $\beta_0 < 0.73$~arcsec. The most probable seeing value is 0.82~arcsec.
\item The free atmosphere median seeing $\beta_{free}$ is 0.51~arcsec, the mode is 0.35~arcsec.
\item The best seeing (minimal OT strength) is observed in October -- November. The typical median value for that period is $\approx 0.83$~arcsec.
\item In conditions of the seeing better than a median one the optical turbulence is dominated by the ground layer, its fraction reaches 0.65.
\item The periods of bad image quality occur when OT in the boundary and free atmosphere intensifies. 
\item The median isoplanatic angle is 2.07~arcsec in general and 2.38~arcsec in conditions of the seeing better than the median one.
\item The median corrected atmospheric time constant is $2.58$\,ms, it exceeds $3.3$\,ms in conditions of weak turbulence.
\end{itemize}

These results are based on only two years monitoring and reflect the current situation at the site under evaluation. It is well known that the astroclimate parameters are subject of long-term trends and cycles, so the continued monitoring may introduce certain corrections. That is why we proceed further with the ASM
operation. When the 2.5-m telescope will be put into operation at the observatory, the ASM functionality will be involved in the flexible observations planning.

The more extended site astronomical characterization demands also the monitoring of atmospheric extinction (also in the IR domain), of the sky background level and of such an important characteristic for AO as an altitude wind speed profile.

Nevertheless, it is already possible to characterize the potential of the studied summit as moderately high for astronomical observations. The basic parameters are slightly worse with respect to those of leading astronomical observatories, but not very significantly. The main conclusion is that modern methods of improving telescope resolution will efficient at Shatdzhatmaz.

\section{Acknowledgments}

The reported site testing and OT properties research at Mt.~Shatdzhatmaz were performed under the partial support of the RFBR (grant 06-02-16902Á). Authors are grateful to the Pulkovo solar station staff, Sternberg institute colleagues P.\,Kortunov, A.\,Belinsky, M.\,Kuznetsov and E.\,Gorbovskoy and students of the astronomical department of the Moscow University physical faculty for their help in the work. A special thank is to A.\,Tokovinin for numerous advices and discussions on the OT research problems throughout the project.

\label{lastpage}


\begin{thebibliography}{99}
\bibitem[\protect\citeauthoryear{Avila et al.}{2008}]{lolas08} Avila R., Avil\'es J.L., Wilson R.W., Chun M., Butterley T., Carrasco E., 2008, MNRAS, 387, 1511

\bibitem[\protect\citeauthoryear{Boshnjakovitch et al.}{1962}]{Boshnjak62} Boshnjakovitch N.A., Bystrova N.V., Demidova A.N., Zhukova L.N., Kassinsky V.V., Eerme K.A., 1962, Izvestiia GAO v Pulkove, 22, Vyp (No) 5, 155

\bibitem[\protect\citeauthoryear{Bowen}{1964}]{bowen1964} Bowen, I. S., 1964, AJ, 69, 816

\bibitem[\protect\citeauthoryear{Carrasco, Avila \& Carrami\~nana}{Carrasco et al.}{2005}]{Carrasco} Carrasco E., Avila R., Carrami\~nana A., 2005,
PASP, 117, 104

\bibitem[\protect\citeauthoryear{Darchiya, Schmeel \& Darchiya}{Darchiya et al.}{1960}]{Darchija60} Darchiya A.H., Schmeel L.F., Darchiya S. P., 1960, Izvestiia GAO v Pulkove, 21, Vyp. (No) 6, 52

\bibitem[\protect\citeauthoryear{Depenchuk, Kondratyuk \& Kojfman}{Depenchuk et al.}{1977}]{Ter1} 	Depenchuk E.A., Kondratyuk R.R., Kojfman A.P., 1977, in Astrometr. Astrofiz., Vyp. (No.) 31, Naukova dumka, Kiev, p. 99

\bibitem[\protect\citeauthoryear{Ehgamberdiev et al.}{2000}]{eso} Ehgamberdiev S. A., Baijumanov A. K., Ilyasov S. P., et al., 2000, A\&A Suppl., 145, 293

\bibitem[\protect\citeauthoryear{Els et al.}{2009}]{ctio4} Els S.G., Sch\"ock M., Bustos E., Seguel J., Vasquez J., Walker D., Riddle R., Skidmore W., Travouillon T., Vogiatzis K.,
2009, PASP, 121, 922

\bibitem[\protect\citeauthoryear{Fuchs, Tallon \& Vernin}{Fuchs et al.}{1998}]{Fuchs} Fuchs A., Tallon M., Vernin J., 1998, PASP, 110, 86

\bibitem[\protect\citeauthoryear{Fuensalida et al.}{2004}]{TPcan} Fuensalida J., Chueca S., Delgado J.M. et al, 2004,
in Ardeberg A.L., Andersen T., eds, Second Backaskog Workshop on Extremely Large Telescopes, Proc. SPIE 5382, p. 619

\bibitem[\protect\citeauthoryear{Gnevysheva}{1991}]{gnevkg} Gnevysheva K.G., 1991, Izvestiia GAO v Pulkove, 207, 77

\bibitem[\protect\citeauthoryear{Hardy}{1998}]{Hardy} Hardy J.W., 1998, Adaptive Optics for Astronomical Telescopes, Oxford University Press

\bibitem[\protect\citeauthoryear{Kornilov \& Kornilov}{2010}]{exp2010} Kornilov V., Kornilov M., 2010,\\ preprint (arXiv:1005.4658v1 [astro-ph.IM])

%\bibitem[\protect\citeauthoryear{Kornilov \& Safonov}{2010}]{dm2010} Kornilov V., Safonov B., 2010, Specific features of differential motion in the DIMM, http://curl.sai.msu.ru/\\mass/download/doc/dimm\_specs.pdf

\bibitem[\protect\citeauthoryear{Kornilov et al.}{2003}]{MASS} Kornilov V., Tokovinin A., Voziakova O., Zaitsev A., Shatsky N., Potanin S., Sarazin M., 2003, in Wizinovich P.L., Bonaccini D., eds, Adaptive Optical System Technologies II. Proc. SPIE 4839, p. 837

\bibitem[\protect\citeauthoryear{Kornilov et al.}{2007}]{MD2007} Kornilov V., Tokovinin A., Shatsky N., Voziakova O., Potanin S., Safonov B., 2007, MNRAS, 382, 1268

\bibitem[\protect\citeauthoryear{Kornilov et al.}{2009}]{maid2005} Kornilov V., Ilyasov S., Vozyakova O., Tillaev Yu., Safonov B., Ibragimov M., Shatsky N., Egamberdiev Sh., 2009, Astronomy Letters, 35, 547

\bibitem[\protect\citeauthoryear{Leushin et al.}{1975}]{SAO2} Leushin V.V., Nelyubin N.F., Nebelitskij V.B., Novikov S.B.
1975, Astron. Tsirk, No. 866, 6

\bibitem[\protect\citeauthoryear{Lipunov et al.}{2010}]{master} Lipunov V., Kornilov V., Gorbovskoy E., Shatskij N. et al., 2010, Advances in Astronomy, 2010, 1

\bibitem[\protect\citeauthoryear{Martin}{1987}]{Martin1987} Martin H.M., 1987, PASP, 99, 1360

\bibitem[\protect\citeauthoryear{Nelyubin \& Vasilyev}{1969}]{SAO1} Nelyubin N.F., Vasilyev O. B., 1969
Astrofiz. Issled. Izv. Spets. Astrofiz. Obs., 1, 185

%\bibitem[\protect\citeauthoryear{Rajagopal et al.}{2008}]{lusci} Rajagopal J., Tokovinin A., Bustos E., Sebag J., 2008, in Scholler M., Danchi W. C., Delplancke F., eds, Optical and Infrared Interferometry, Proc. SPIE 7013, p. 1P

\bibitem[\protect\citeauthoryear{Riddle, Schoeck \& Skidmore}{Riddle et al.}{2006}]{TMTRobot} Riddle R., Schoeck M., Skidmore W.,
2006, in Stepp L., ed., Ground-based and Airborne Telescopes, Proc. SPIE 6267, p. 1Q

\bibitem[\protect\citeauthoryear{Roddier}{1981}]{Roddier81} Roddier F., 1981, in Wolf E., ed., Progress in Optics, 19, North-Holland, Amsterdam, p. 281

\bibitem[\protect\citeauthoryear{Roddier}{1999}]{Rodier_book} Roddier F., 1999, Adaptive optics in astronomy Cambridge, Cambridge Univ. Press, New York

\bibitem[\protect\citeauthoryear{Rylkov \& Bobylev}{1991}]{Rylkov} Rylkov V.P., Bobylev V.V., 1991, Izvestiia GAO v Pulkove, 207, 89

\bibitem[\protect\citeauthoryear{Sarazin \& Roddier}{1990}]{DIMM} Sarazin M., Roddier F., 1990, A\&A. 227, 294

\bibitem[\protect\citeauthoryear{Saunders et al.}{2009}]{DomeC} Saunders W., Lawrence J., Storey J. et al, 2009, PASP, 121, 976

\bibitem[\protect\citeauthoryear{Schoeck et al.}{2009}]{TMT} Schoeck M., Els S., Riddle R. et al., 2009, PASP, 121, 384

\bibitem[\protect\citeauthoryear{Skidmore et al.}{2009}]{tmtV} Skidmore W., Els S., Travouillon T., Riddle R., Sch\"ock M., Bustos E., Seguel J., Walker D.,
2009, PASP, 121, 1151

\bibitem[\protect\citeauthoryear{Soules et al.}{1996}]{Soules1996} Soules D.B., Drexler J.J., Draayer B.F., Eaton F.D., Hines J.R., 1996, PASP, 108, 817

\bibitem[\protect\citeauthoryear{Tatarskii}{1967}]{Ta67} Tatarskii V.I., 1967, Wave Propagation in a Turbulent Medium. Dover Press, New York

\bibitem[\protect\citeauthoryear{Tokovinin}{2002a}]{T2002} Tokovinin A., 2002a, PASP, 114, 1156

\bibitem[\protect\citeauthoryear{Tokovinin}{2002b}]{T2002time} Tokovinin A., 2002b, Appl. Opt., 41, 957
 
\bibitem[\protect\citeauthoryear{Tokovinin, Baumont \& Vasquez}{Tokovinin et al.}{2003}]{CTIO03} Tokovinin A., Baumont, S., Vasquez J., 2003, MNRAS, 340, 52

\bibitem[\protect\citeauthoryear{Tokovinin \& Kornilov}{2007}]{TK07}
Tokovinin A., Kornilov V., 2007, MNRAS, 381, 1179

\bibitem[\protect\citeauthoryear{Tokovinin, Sarazin \& Smette}{2007}]{VLT07} Tokovinin A., Sarazin M., Smette A., 2007, MNRAS, 378, 701

\bibitem[\protect\citeauthoryear{Tokovinin \& Travouillon}{2006}]{CP2006} Tokovinin A., Travouillon T., 2006, MNRAS, 365, 1235

\bibitem[\protect\citeauthoryear{Tokovinin et al.}{2003}]{mnras2003} Tokovinin A., Kornilov V., Shatsky N., Voziakova O., 2003, MNRAS, 343, 891 (TKSV)

\bibitem[\protect\citeauthoryear{Tokovinin et al.}{2010}]{lusci} Tokovinin A., Bustos E., Berdja A., 2010,
MNRAS, 404, 1186

\bibitem[\protect\citeauthoryear{Thomas-Osip et al.}{2008}]{gmt2008} Thomas-Osip J. E., Prieto G., Johns M., Phillips M.M., 2008, in Stepp L.M., Gilmozzi R., eds, Ground-based and Airborne Telescopes II, Proc. SPIE 7012, p. 1U

\bibitem[\protect\citeauthoryear{Thomsen, Britton \& Pickles}{Thomsen et al.}{2007}]{Palomar} Thomsen M., Britton M., Pickles A., 2007, BAAS, 38, 242

\bibitem[\protect\citeauthoryear{Travouillon et al.}{2008}]{tmt-temp} Travouillon T., Els S.G., Riddle R.L., et al., 2008,
in Stepp L.M., Gilmozzi R., eds, Ground-based and Airborne Telescopes II, Proc. SPIE 7012, 20

\bibitem[\protect\citeauthoryear{Travouillon et al.}{2009}]{tmtVII} Travouillon T., Els S., Riddle R. L., Sch\"ock M., Skidmore W., 2009, PASP, 121, 787

%\bibitem[\protect\citeauthoryear{Vasilyev \& Vjazovov}{1962}]{Vasiljev62} Vasilyev O.B., Vjazovov V.V., 1962, Izvestiia GAO v Pulkove, 22, Vyp.5, 144

\bibitem[\protect\citeauthoryear{Vernin et al.}{1991}]{vern1991} Vernin J., Caccia J.-L., Weigelt G., Mueller M., 1991, A\&A, 243, 553

\bibitem[\protect\citeauthoryear{Vernin, Mu\~noz-Tu\~non \& Sarazin}{Vernin et al.}{2008}]{EELT} Vernin J., Mu\~noz-Tu\~non C., Sarazin M., 2008, in L. Stepp, R. Gilmozzi, eds, Ground-based and Airborne Telescopes II, Proc. SPIE 7012, p. 1T

\bibitem[\protect\citeauthoryear{Wilson}{2002}]{wilson02} Wilson R.W., 2002, MNRAS, 337, 103

\bibitem[\protect\citeauthoryear{Wilson et al.}{2003}]{Wils2003} Wilson R. W., Wooder N.J., Rigal F., Dainty J. C., 2003, MNRAS, 339, 491

\bibitem[\protect\citeauthoryear{Wilson, Butterley \& Sarazin}{Wilson et al.}{2009}]{Paranal} Wilson R., Butterley T., Sarazin M., 2009, MNRAS, 399, 2129

\end{thebibliography}
\end{document}